\documentclass[conference]{IEEEtran}
\IEEEoverridecommandlockouts
\usepackage{cite}
\usepackage{amsmath,amssymb,amsfonts}
\usepackage{algorithmic}
\usepackage{graphicx}
\usepackage{textcomp}
\usepackage{xcolor}
\usepackage{url}
\usepackage{subfigure}
\usepackage{multirow}
\usepackage{comment}
\usepackage{colortbl}
\usepackage{hhline}
\def\BibTeX{{\rm B\kern-.05em{\sc i\kern-.025em b}\kern-.08em
    T\kern-.1667em\lower.7ex\hbox{E}\kern-.125emX}}
\begin{document}

\title{EHAAS: Energy Harvesters As A Sensor for Place Recognition on Wearables\\
}

\author{\IEEEauthorblockN{Yoshinori Umetsu$^1$,
Yugo Nakamura$^1{}^,{}^2$,
Yutaka Arakawa$^1{}^,{}^3$,
Manato Fujimoto$^1$,
Hirohiko Suwa$^1$
}
\IEEEauthorblockA{$^1$
Graduate School of Information Science, 
Nara Institute of Science and Technology, 
Ikoma, Nara 630-0192, JAPAN\\
E-mail: \{umetsu.yoshinori.uv4, Nakamura.yugo.ns0, ara, manato, h-suwa\}@is.naist.jp\\
$^2$JSPS Research Fellowships for Young Scientists\\
$^3$JST PRESTO}
}

\maketitle

\begin{abstract}
A wearable based long-term lifelogging system is desirable for the purpose of reviewing and improving users’ lifestyle habits. Energy harvesting (EH) is a promising means for realizing sustainable lifelogging. However, present EH technologies suffer from instability of the generated electricity caused by changes of environment, e.g., the output of a solar cell varies based on its material, light intensity, and light wavelength. In this paper, we leverage this instability of EH technologies for other purposes, in addition to its use as an energy source. Specifically, we propose to determine the variation of generated electricity as a sensor for recognizing ``places'' where the user visits, which is important information in the lifelogging system. First, we investigate the amount of generated electricity of selected energy harvesting elements in various environments. Second, we design a system called EHAAS (Energy Harvesters As A Sensor) where energy harvesting elements are used as a sensor. With EHAAS, we propose a place recognition method based on machine-learning and implement a prototype wearable system. Our prototype evaluation confirms that EHAAS achieves a place recognition accuracy of 88.5\% F-value for nine different indoor and outdoor places. This result is better than the results of existing sensors (3-axis accelerometer and brightness). We also clarify that only two types of solar cells are required for recognizing a place with 86.2\% accuracy. 
\end{abstract}


\section{Introduction}
Reviewing and improving people’s lifestyle habits is essential to treat lifestyle-related diseases such as diabetes, hypertension, and mental health issues. Lifelogging system records the personal data generated by a user’s behavioral activities. It is attracting attention for the purpose of understanding life patterns and/or lifestyle habits. Nowadays, a lifelog is recorded automatically by a smartphone or wearable activity trackers. For example, the mobile application ``Moves''\footnote{Moves - Activity Diary for iPhone and Android : \\\url{https://moves-app.com/}} can track location, activities, and transportation. Sony’s lifelogging application ``Lifelog''\footnote{Lifelog - innovative activity tracker Android app from Sony : \\\url{https://www.sonymobile.com/global-en/apps-services/lifelog/}} tracks which applications are used and what their usage time is. It also logs music played, photos taken, etc. An activity tracker adds more detailed information such as steps, heart rate, and sleep status. 

One piece of critical information in the lifelog is location. Recording an accurate position is not essential, but a room-level location (which we call \emph{place}, hereafter) is required for assessing a user’s lifestyle habits or life pattern. Basically, the aforementioned applications including Moves, record location using GPS embedded in a smartphone. GPS is ubiquitous but it does not work indoors\cite{kourogi2006indoor}. Therefore, it can only provide a building-level location. To record room-level indoor places such as an office room, meeting room, lecture room, or toilet, other localization technologies are required.

For indoor places, WiFi-based localization\cite{6129175,6042868,WOO20113,cypriani2009open,ciurana2007wlan} is the most popular because WiFi access points have already been deployed widely and everyone has a WiFi-enabled smartphone. However, from the viewpoint of long-term lifelogging for lifestyle improvement, smartphone-based approaches are not suitable. The current mobile operating systems do not allow applications to keep tracking the location in the background because of battery conservation. Active/passive RFID-based indoor localization systems\cite{4812905,Xiao:2018:OMT:3190721.3190729} are also popular and able to achieve cm-level accuracy. However, they need some infrastructure (RFID tags/RFID readers) in environments and continuous/regular power charging of tags/readers for daily sensing. For our purpose, iBeacon\footnote{iBeacon : \\\url{https://developer.apple.com/ibeacon/}} from Apple, where iOS keeps monitoring the Bluetooth signals in the operating system, might be the best solution. However, it is costly because it requires the deployment of new beacon tags into each place and the perpetual maintenance of their batteries. No wearables (including Apple watch) can keep sensing iBeacon signals continuously because that would drastically deplete the small battery of a wearable.

In this paper, we propose a wearable lifelogging system called EHAAS (Energy Harvesters As A Sensor), which utilizes a set of different energy harvesting elements as a sensor for recognizing places. Energy Harvesting (EH) has been expected to be used as a power source for running sensors and processors\cite{paradiso2005energy,bharatula2004towards}. It became a reality because the power consumption of those devices becomes smaller every year. In April 2018, a smartwatch called PowerWatch\footnote{Matrix PowerWatch :\\\ \url{https://www.powerwatch.com/}}, which uses thermoelectric EH, was released. It can count the number of steps by a low power accelerometer. However, the generated electricity is not enough to run other sensors or a network module for localization. Therefore, we believe that a fundamentally different approach is necessary to realize room-level localization with wearables.

Our key idea is to leverage a disadvantage of EH that the amount of generated electricity depends on the surrounding environment. Generated electricity of a solar panel varies based on the light conditions such as brightness, wavelength, and angle of the light source. Also, the generated electricity changes when a different material is used for a solar panel. Kinetic EH also has a disadvantage in that it only generates power when it moves physically. Leveraging the instability of EH elements as a feature of places, the variation of generated electricity could be used as a sensor for place recognition. If EH works as a sensor as well as a power supply, there will be a big potential for realizing a place recognition system with lower power consumption. The contributions of this paper are summarized as follows:

\begin{itemize}
 \item \textcolor{black}{First, to investigate the feasibility of our idea, we evaluate the characteristics of various EH devices such as solar cells, peltier elements, and piezo elements. According to previous research, the characteristics of EH devices have not been studied well enough to evaluate the possibility of a sensor to recognize places. We show that EH devices are not only substitute for existing sensors, but can also be used as a sensor that is suitable for place recognition through the characteristic evaluation of EH devices. }
 \item \textcolor{black}{Second, we evaluate the amount of generated electricity in various places to establish a relationship between places and generated electricity in those places. The target places in this paper comprise common places in our university such as a classroom and a restroom, aiming to apply the developed technology to an office environment. We show that each place generates a unique amount of electricity according to its environment or the movement of a user, and EH devices have a large potential for recognizing places.}
 \item \textcolor{black}{Finally, we implement a prototype of a wearable device combining multiple EH elements as a sensor. Through experiments with the prototype, we confirm that EHAAS has achieved a place recognition accuracy of 88.5\% F-value for nine different places including indoors and outdoors in the university. Additionally, we compare this result with the accuracy recognized by existing sensors such as accelerometers and luminometers. As a result, a higher accuracy is obtained with EHAAS. Furthermore, we clarify that this system with only two types of solar cells can recognize a place with 86.2\% accuracy.}
\end{itemize}


The rest of this paper is organized as follows. Section \ref{Related work} reviews existing work and energy harvesting elements related to this paper. The preliminary measurements are provided in Section \ref{preliminary measurement}, and Section \ref{place recognition} presents our proposed place recognition system with energy harvesters. In Section \ref{performance evaluation}, we describe the performance evaluation and evaluation results. Finally, Section \ref{conclusion} concludes this paper.

\section{Related work}
\label{Related work}
As related work, we summarize existing studies of both indoor localization techniques and energy harvesting technologies. 

\subsection{Indoor localization}
Indoor localization techniques\cite{DBLP:journals/corr/abs-1709-01015} have been widely studied. Sensor data and algorithms used for localization are diverse, depending on the target accuracy\cite{mautz2012indoor}. In this paper, we only consider the techniques that are feasible on ordinary devices and are for room-level localization (the target of this paper)\cite{kyritsis2016ble}. Examples include a radio-based technique that utilizes WiFi or Bluetooth, a dead reckoning with inertial sensors\cite{pai2012padati}, and acoustic fingerprinting\cite{tarzia2011indoor}. 

WiFi-based localization methods usually adopt the technique called fingerprinting\cite{brunato2005statistical}. This technique surveys a fingerprint of radio signals at various places in advance. Based on the similarity of fingerprints, it estimates the location. In the case of BLE beacons\cite {zhuang2016smartphone,kyritsis2016ble}, the method is basically the same, and Kyritsis A.I., et al.\cite{kyritsis2016ble} achieved 81.17\% $\sim$ 100\% accuracy (90.2\% on average) with fingerprints at 63 different places in seven rooms. However, it is difficult to continuously scan WiFi or BLE signals for lifelogging on wearable devices that only have a small battery. In fact, most of the studies such as \cite{kyritsis2016ble} use a laptop or a smartphone, not a smartwatch, for the experiment. 

As a study of localization using a smartwatch, Lee et al. have proposed a place identification method for a home environment\cite{lee2015non} that utilizes activity fingerprinting obtained by a motion sensor. However, since this research is based on the premise that the movement of an arm is different for each room, it cannot identify when the same work (PC or meeting) is done in different rooms, like in a university or office. Also, similar to the case with radio signal scanning, continuous use of the acceleration sensor consumes battery power and is not realistic for long-term lifelogging by a wearable device.

\subsection{Energy harvesting elements}
Below, we introduce representative energy harvesting elements.

\noindent
\textbf{Photovoltaic (Solar cell): }
Solar cells, which use the photovoltaic effect, are the most popular energy harvesting elements because they are inexpensive and have high flexibility in size. There are several kinds of materials having a photovoltaic effect. The amount of generated electricity depends on the kind of material in addition to the brightness and wavelength. Representative materials are amorphous silicon, polycrystalline silicon, and organic thin films. Solar cells are widely used in calculators and traditional watches. Sometimes, they are attached to bags and backpacks. However, they are not used for the latest wearable devices because of the need for a small cell size. We can expect that, in the future, they will be embedded into clothes because RIKEN has released a novel solar cell that is launderable as well as pasteable using an iron.

\noindent
\textbf{Kinetic (Piezo element): }
A piezoelectric element utilizes the piezo effect that generates electricity by distortion. As an example of use as a power supply, the power generating floor produced by Soundpower Corporation\footnote{Soundpower corporation :\\ \url{http://www.soundpower.co.jp/}} in Japan is well-known. However, this element is usually used as a pressure sensor. We used it for sensing the posture of a worker on a chair\cite{otoda2018census-backup} and for sensing the weight distribution on a glove\cite{akpa2018perhealth-backup}. H. Kalantarian et al. have used a piezo element as an insole pressure sensor and a power source \cite{Kalantarian}. Also, S. Khalifa et al. have used it as a motion sensor and a power source \cite{khalifa2017harke}. However, in all of these research studies, including our glove, electricity is not generated unless it is in force or in motion. Therefore, it cannot be used for recognizing places.

\noindent
\textbf{Thermoelectrics (Peltier element): }
The peltier element is an energy harvester using the peltier effect that generates electricity by a temperature difference. Usually, it is used for cooling a CPU by flowing current to the element. In April 2018, a smartwatch called PowerWatch using a peltier element as a power source was released. This watch is equipped with LCD, BLE, and a step counting function. Since the temperature difference is critical, it is a problem that electricity is not generated when the body temperature and the outside air temperature are similar.

\section{Preliminary measurement}
\label{preliminary measurement}
Prior to designing the place recognition system, we investigate the characteristics of the following energy harvesting elements: solar cells (5 types), piezo element, and peltier element. 

\begin{table*}[t]
\centering
\caption{Investigated solar cells}
\label{solar-list}
\scalebox{0.7}{
{\renewcommand{\arraystretch}{1.2}
\begin{tabular}{p{2cm}|p{4cm}p{4cm}p{4cm}p{4cm}p{4cm}}
\hline\hline
					& \multicolumn{1}{c}{SC1} & \multicolumn{1}{c}{SC2} & \multicolumn{1}{c}{SC3} & \multicolumn{1}{c}{SC4} & \multicolumn{1}{c}{SC5} \\\hline
 &&&&& \\ 
\raisebox{1.5cm}[0pt][0pt]{Picture}     & \multicolumn{1}{c}{\includegraphics[width=2cm]{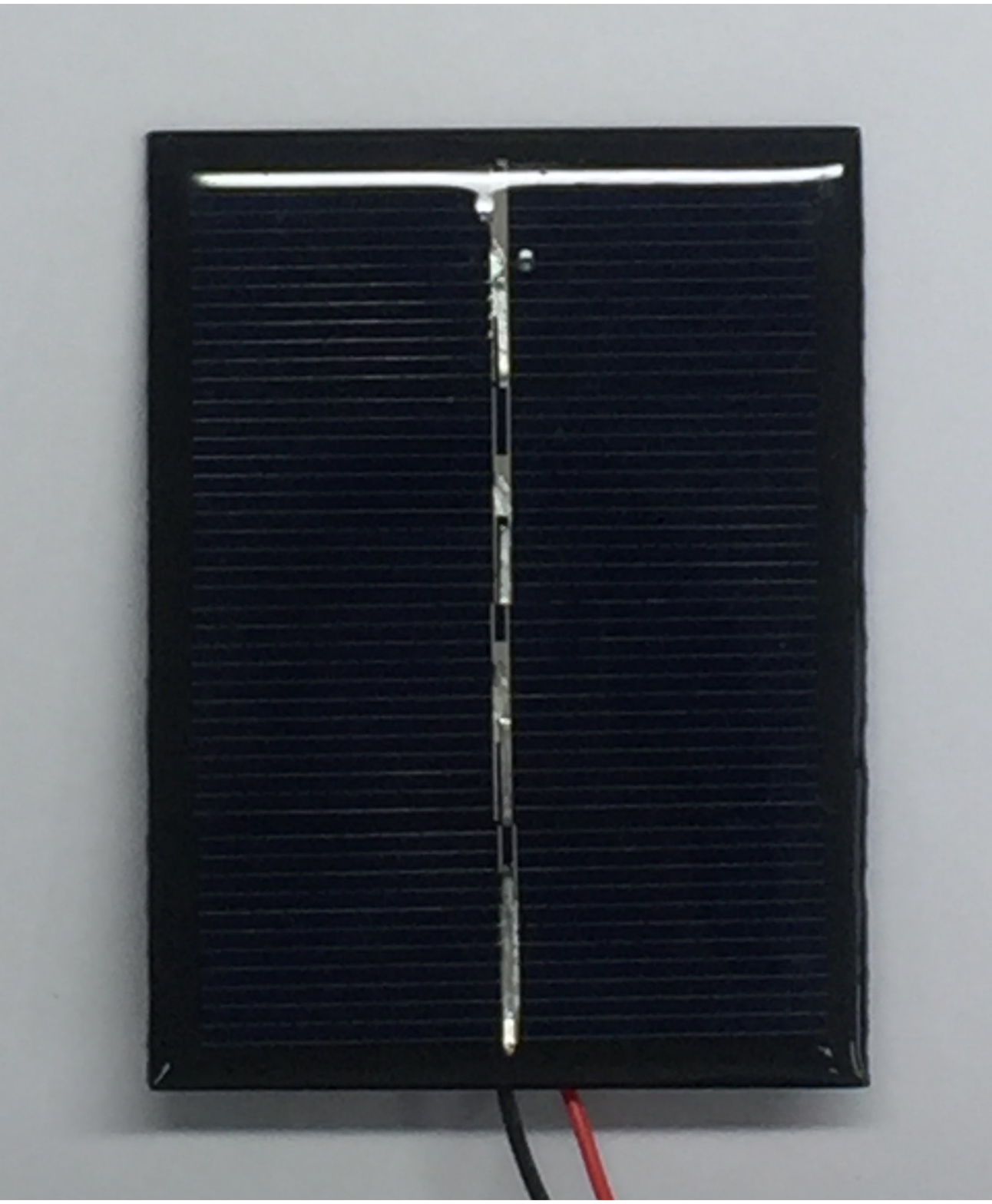}} & \multicolumn{1}{c}{\includegraphics[width=2cm]{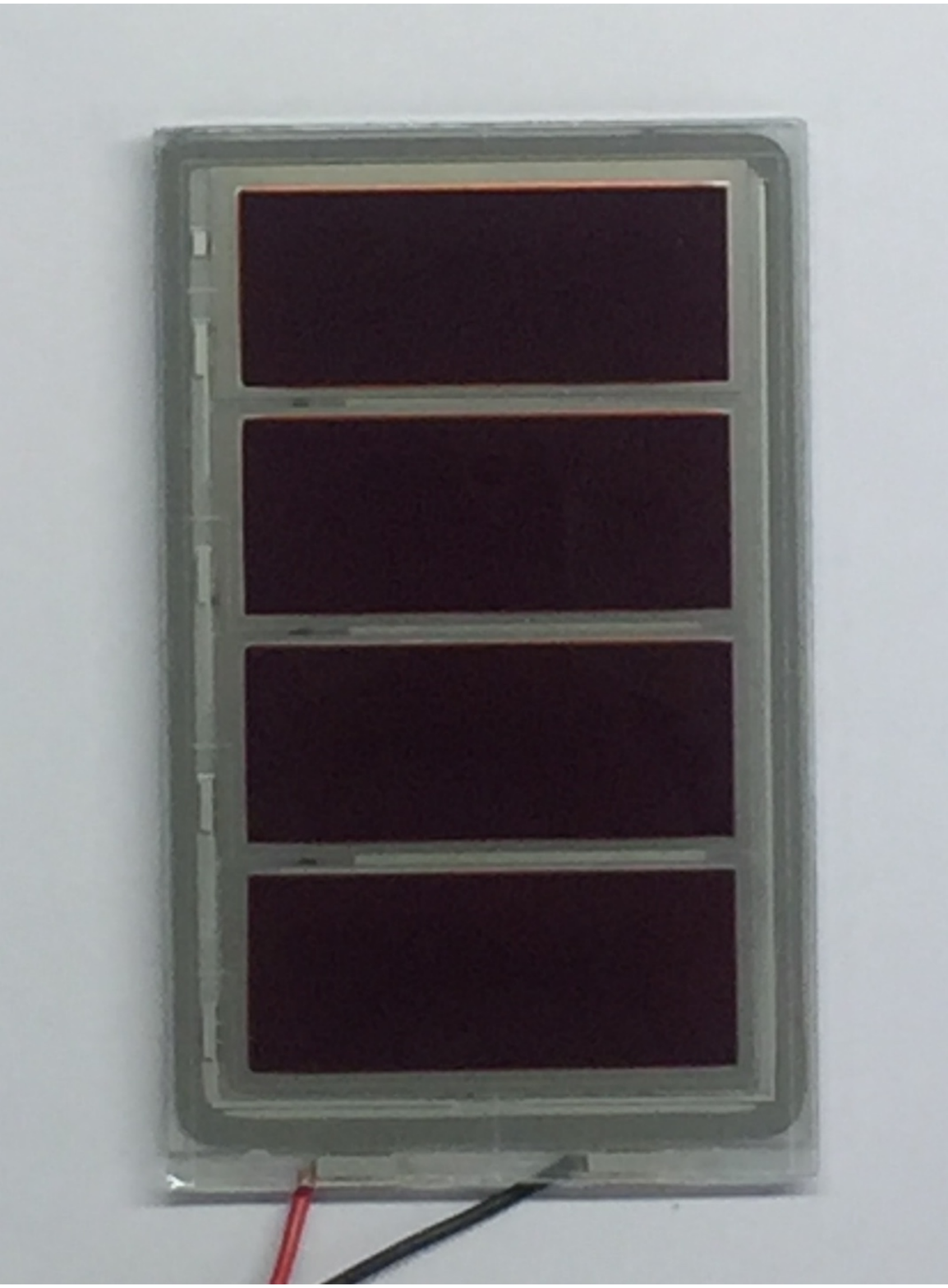}} & \multicolumn{1}{c}{\includegraphics[width=2cm]{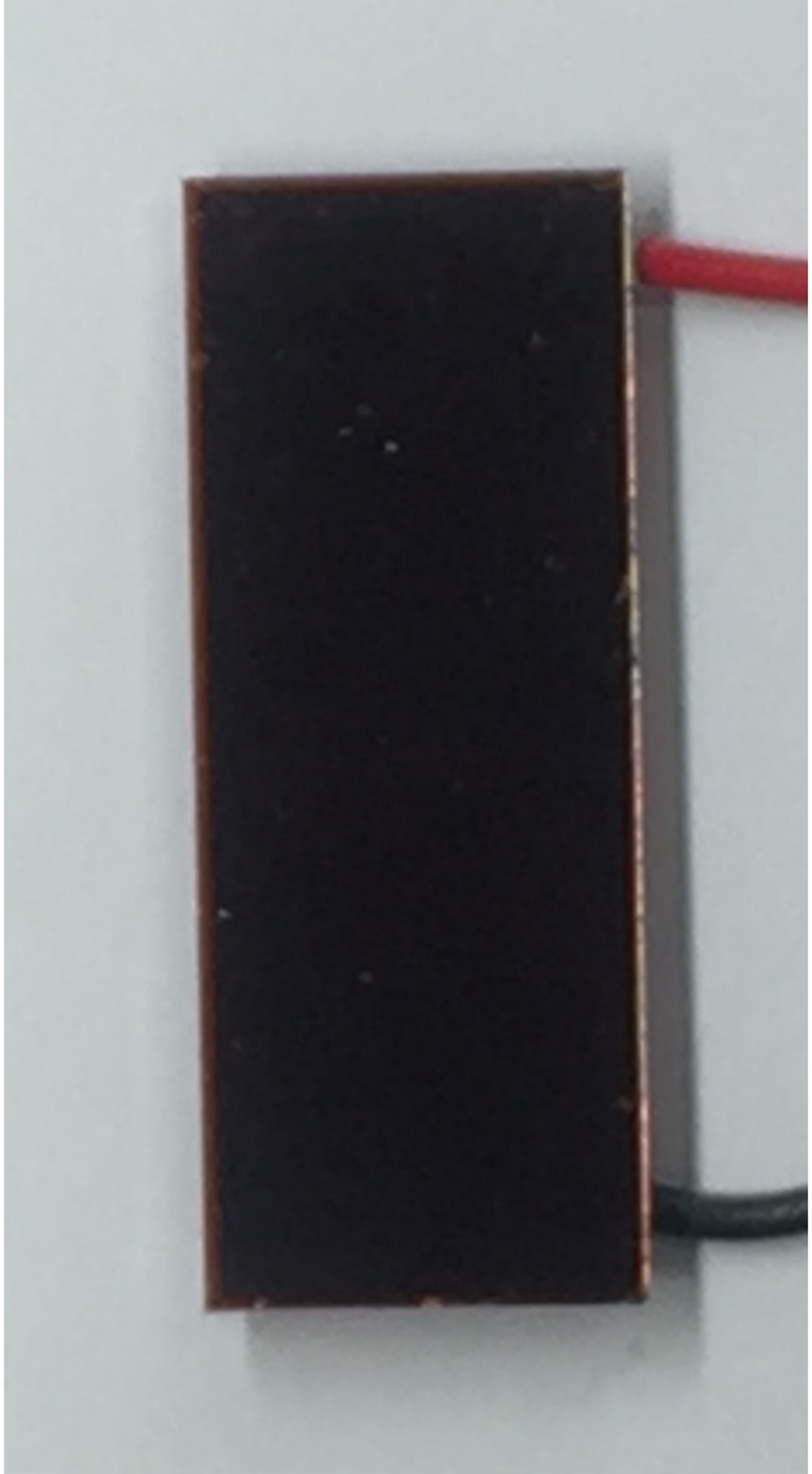}} & \multicolumn{1}{c}{\includegraphics[width=2cm]{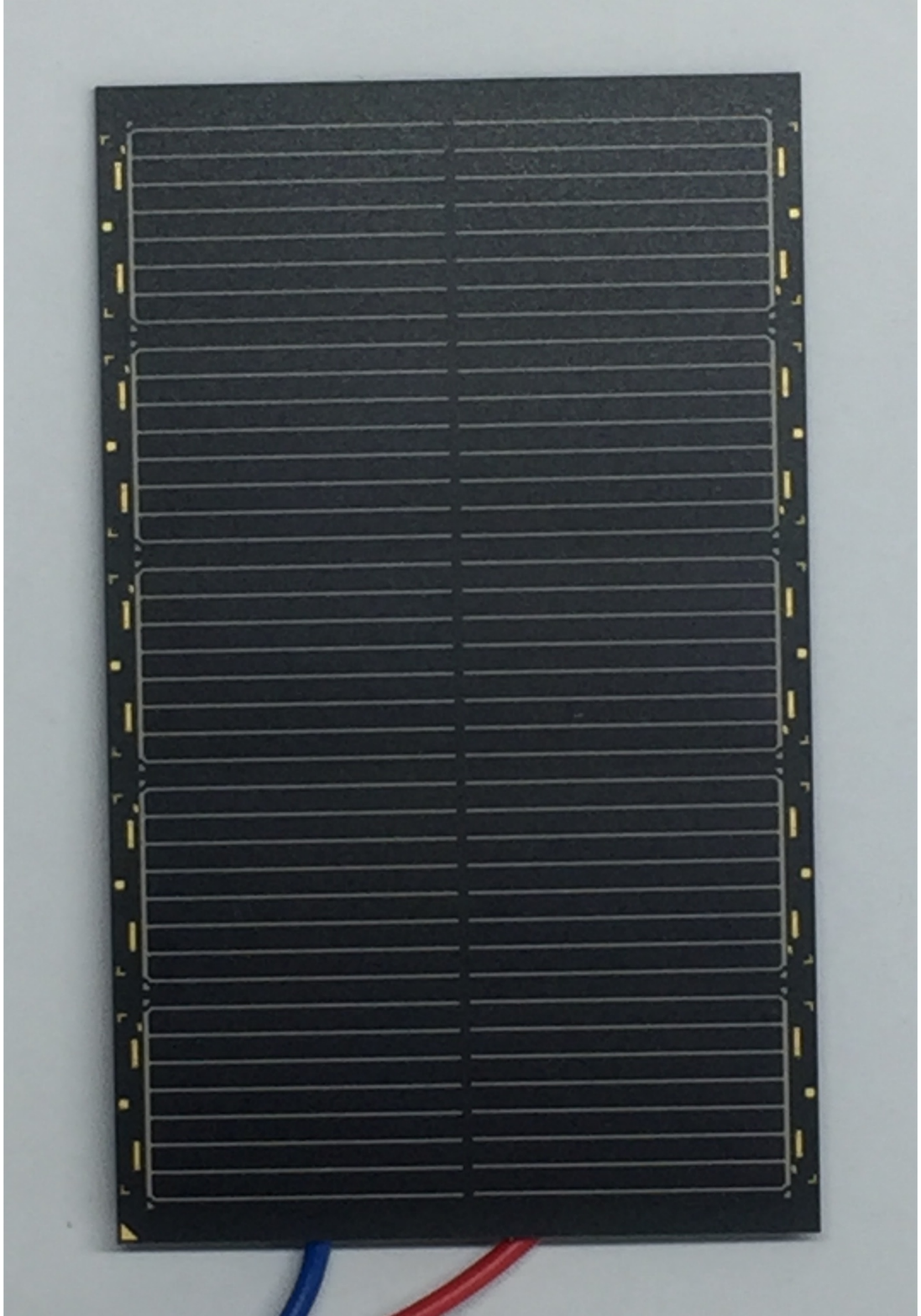}} & \multicolumn{1}{c}{\includegraphics[width=2cm]{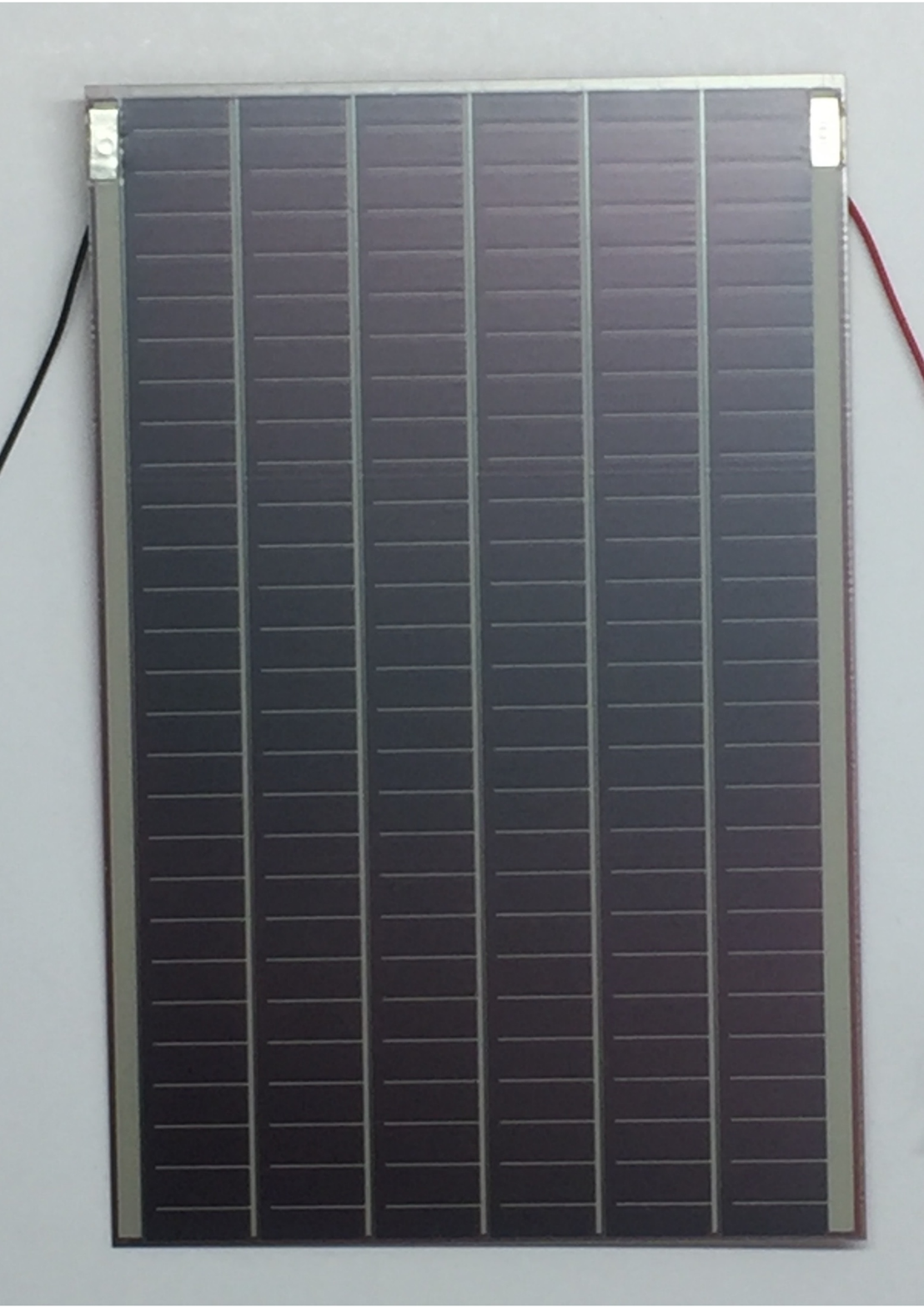}} \\
Manufacture &SUNYOOO solar Limited.&\multicolumn{1}{c}{Fujikura}&\multicolumn{1}{c}{VIMUN}&\multicolumn{1}{c}{SHARP}&\multicolumn{1}{c}{Panasonic}\\
Type        & \multicolumn{1}{c}{Polycrystalline silicon}&\multicolumn{1}{c}{Organic thin film}&\multicolumn{1}{c}{Amorphous silicon}&\multicolumn{1}{c}{Polycrystalline silicon}&\multicolumn{1}{c}{Thin amorphous silicon}\\
Size ($mm^2$)        &\multicolumn{1}{c}{80$\times$60}   &  \multicolumn{1}{c}{91$\times$60}   &  \multicolumn{1}{c}{30$\times$11}  & \multicolumn{1}{c}{67$\times$40}  & \multicolumn{1}{c}{112$\times$73}\\
\hline
\end{tabular}
}
}
\end{table*}

\subsection{Generated-electricity measurement system}
\begin{figure}[t]
\centering
\includegraphics[width=\columnwidth]{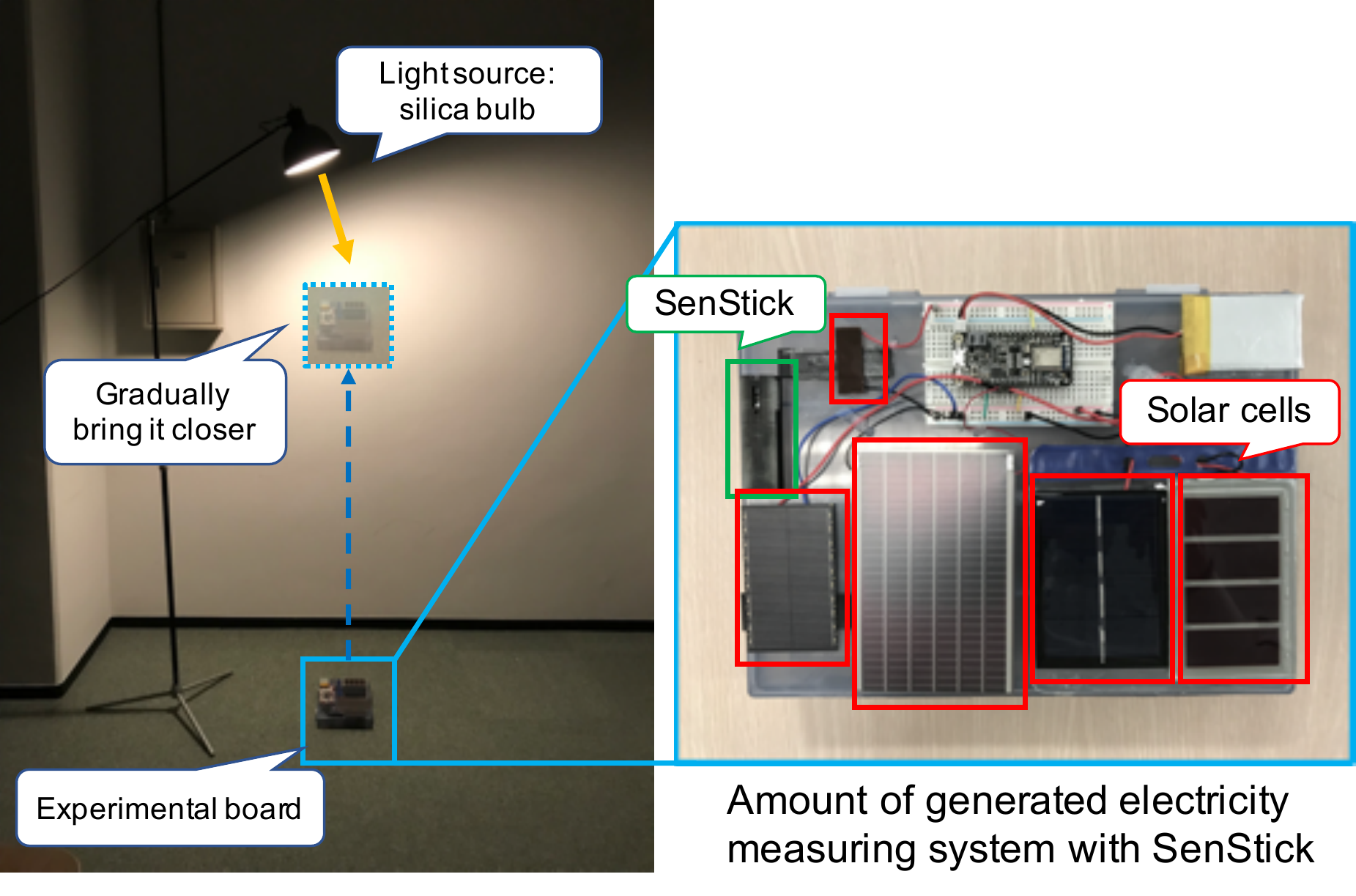}
\caption{Experimental environment for measuring the solar cells}
\label{intensity}
\end{figure}

For the preliminary experiment, we developed a system that can measure the amount of generated electricity of energy harvesters (right figure in Fig. \ref{intensity}).

We utilize an all-in-one Arduino-compatible and Bluetooth Low Energy (BLE) development board (Adafruit Feather nRF 52 Bluefruit LE) that has an ARM Cortex M4F processor clocked at 64MHz and eight analog inputs. Each energy harvesting element is connected to an analog input. The voltage value is acquired as a digital value by AD conversion in the board. The sampling frequency of the voltage value is 50 Hz, and data is transmitted to the laptop via BLE. On the laptop, Python scripts draw graphs and accumulate data. 

In addition, we add SenStick \cite{Nakamura:2017hk-backup} on the experimental board. SenStick is an ultra-tiny multi-sensing board having eight sensors: 9-axis (MPU-9250), temperature/humidity (SHT20), illuminance (BH1780GLI-E2), pressure (LPS25HBTR), and ultraviolet (VEML6070). We use it to collect the data for comparison.

\subsection{Solar cells}
First, we investigate solar cells, which are widely used and easy to buy. There are several types of solar cells with different materials and manufacturing methods, and different characteristics against light conditions such as brightness and wavelength. In this paper, we prepared five small solar cells sold in Japan (Table \ref{solar-list}) and investigated their characteristics. The features of these solar cells are as follows. 

\begin{itemize}
\item SC1, SC4 (Polycrystalline silicon): \\
Polycrystalline silicon is currently the most popular solar cell. SC1 has a glass coating on the surface, while SC4 has no glass coating.

\item SC2 (Organic thin film): \\
Organic thin films are solar cells that can output high voltage even in a low light environment. They increase the light receiving sensitivity to visible light by adsorbing a dye on the surface of titanium oxide. From the viewpoint of manufacturing cost and design ability, it is expected that this solar cell will be widely applied.

\item  SC3, SC5 (Amorphous silicon): \\
Amorphous silicon is a kind of silicon with an irregular crystal structure. Although the energy conversion efficiency is lower than the polycrystalline silicon type solar cell, it can be thinner than polycrystalline silicon and can keep a high voltage output even at high temperature. 
\end{itemize}

\subsubsection{Characteristic against the intensity of the light} 
Fig. \ref{intensity} shows the experimental environment where Panasonic’s silica bulb LW100V 54W was used as a light source in a dark room. We investigated the characteristics against the intensity of the light by changing the distance between the experimental board and the light source.

\begin{figure}[t]
\centering
\includegraphics[width=\columnwidth]{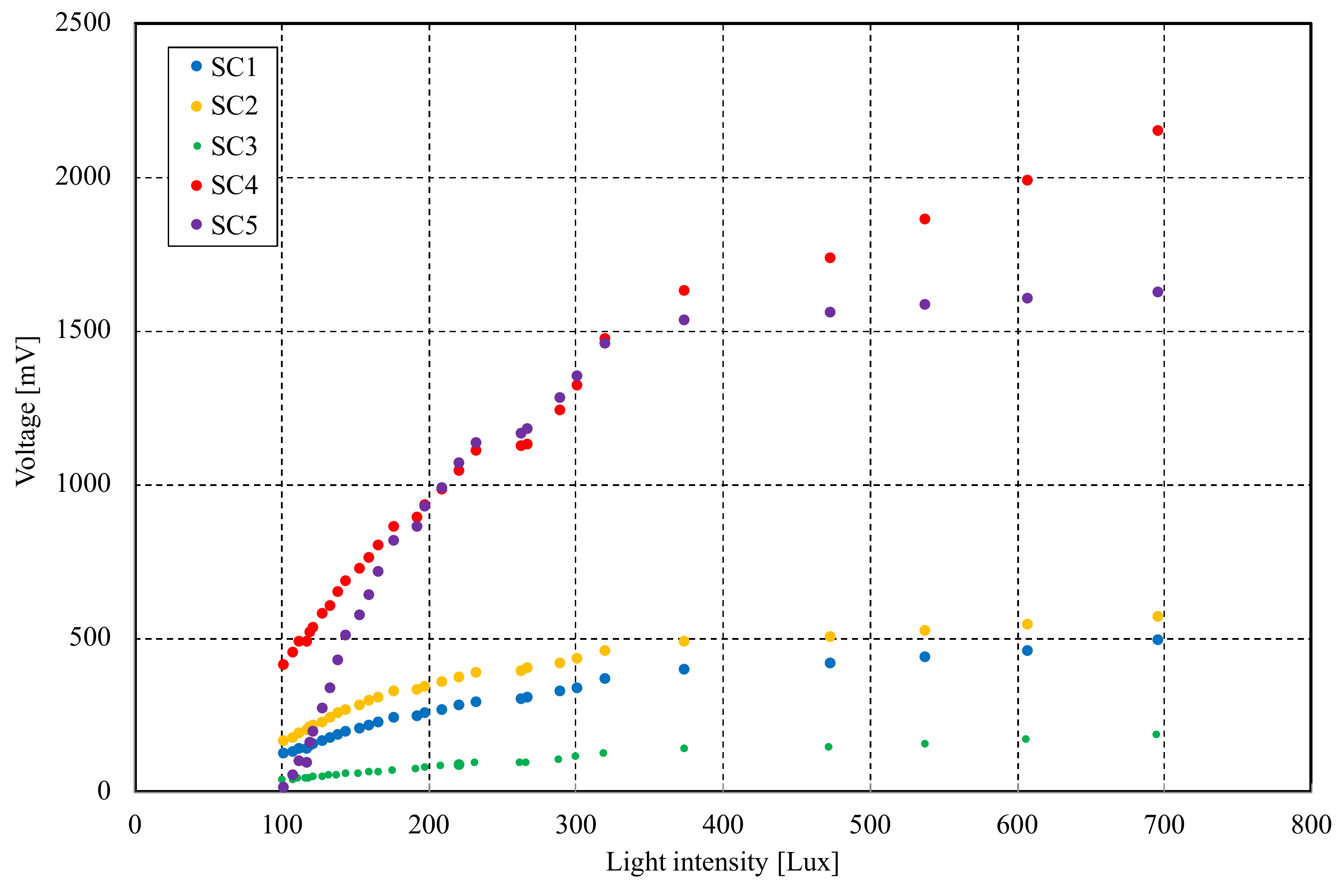}
\caption{Generated electricity against the intensity of the light}
\label{intensity-result-a}
\end{figure}

Fig. \ref{intensity-result-a} shows the amount of generated electricity of each solar cell. The generated electricity of all the solar cells increased linearly according to the light intensity. In particular, SC2 and SC3 generated more electricity than the others. This suggests the possibility that these solar cells can be used as illuminance sensors for distinguishing places. 

\subsubsection{Characteristic against the wavelength of the light} 
Next, we investigated the characteristics against the wavelength of the light. In this experiment, we measure the quantum efficiency (QE), the fraction of photons incident on the solar cell that are converted into electrons without loss, because the energy conversion efficiency of solar cells depends on QE. 

\begin{figure}[t]
\centering
\includegraphics[width=\columnwidth]{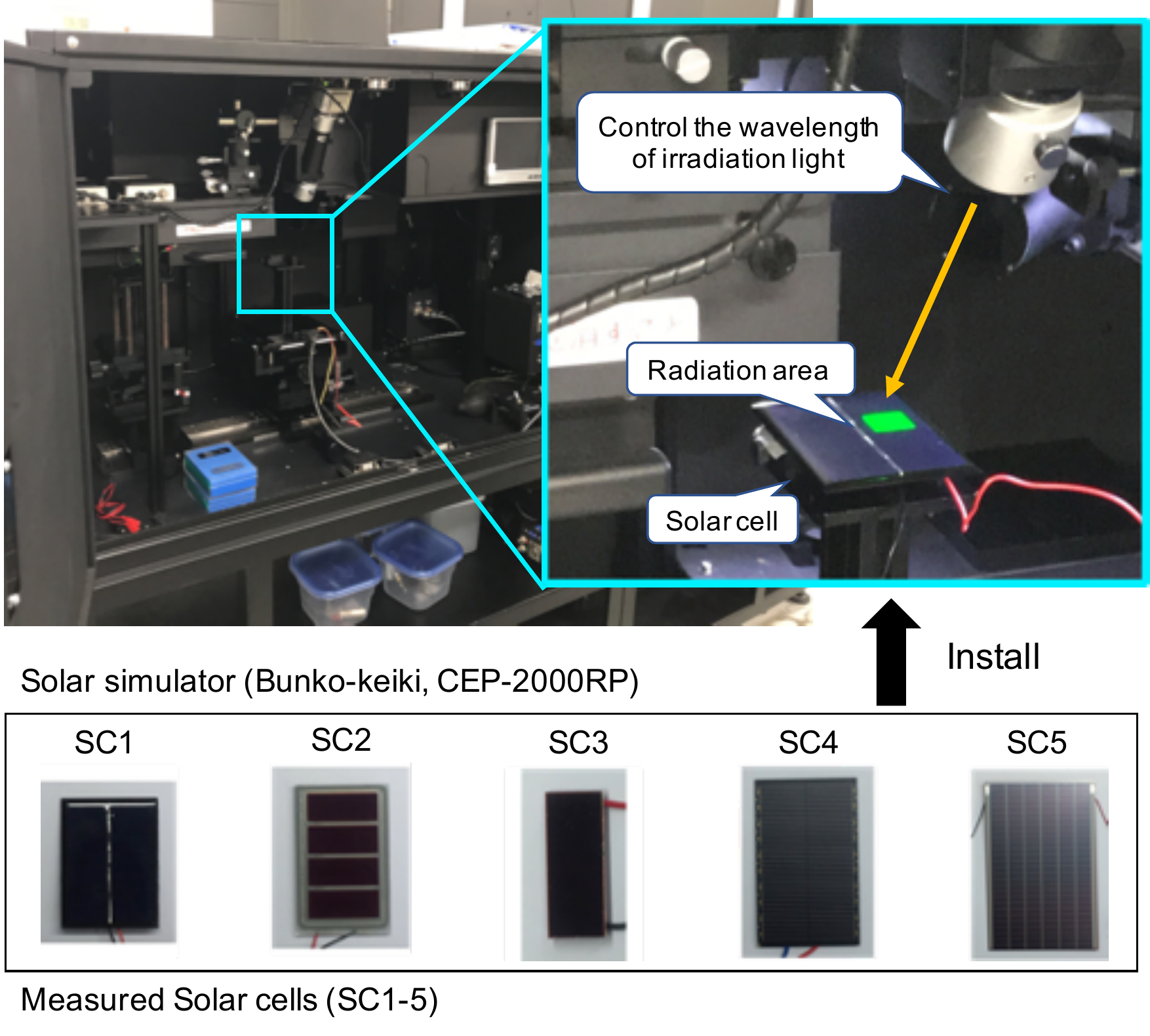}
\caption{A solar simulator for measuring quantum efficiency}
\label{SCs}
\end{figure}

We utilize a solar simulator (Bunko-keiki, CEP-2000RP), shown in Fig. \ref{SCs}, to measure QE under the condition that the intensity of the irradiation light is constant, the wavelength is set to 10 different values in the range 300 nm $ \sim $ 1200 nm. 

\begin{figure}[t]
\centering
\includegraphics[width=.9\columnwidth]{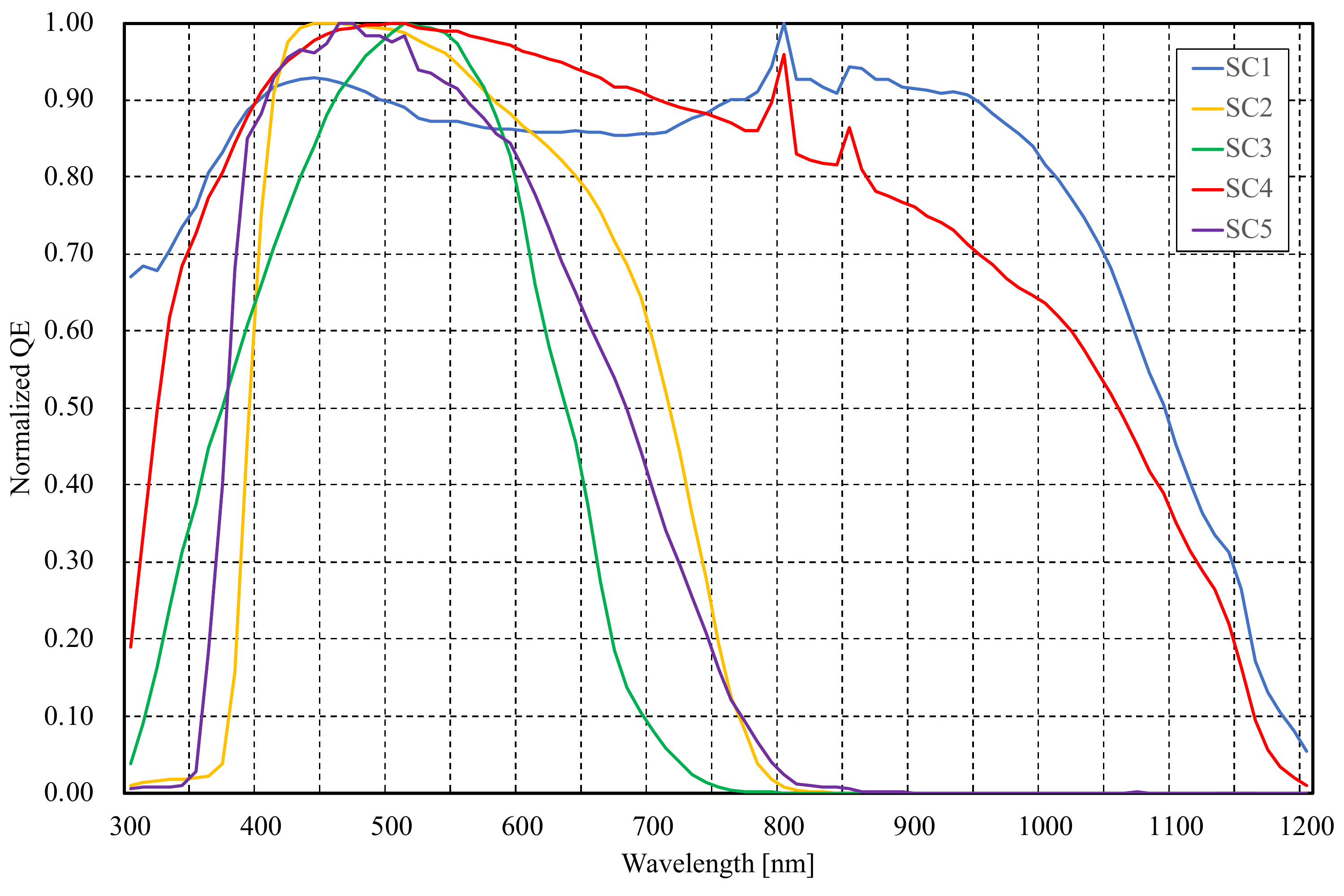}
\caption{Normalized QE against wavelength}
\label{QE}
\end{figure}

Fig. \ref{QE} shows the normalized quantum efficiency (QE) against the wavelength. Small peaks around 800 nm and 850 nm are waveforms generated when the measuring equipment switches light sources, so we ignore them. 

We can observe that the sensitivity against wavelength is different among the types of solar cells. Therefore, this suggests that by utilizing multiple solar cells together, we can get information about the combination of wavelengths at a certain place. At a glance, there is no difference, but in recent offices, different kinds of lighting may often be used in different rooms, according to the purpose of room. This is because the condition of the lighting can affect workers’ mental state, such as emotion\cite{emotion} and fatigue\cite{fatigue}, and optimizing the lighting environment in workplaces is regarded as important\cite{purpose}. Hence, multiple kinds of solar cells may distinguish different places by the characteristic wavelength distribution of each place. 

\subsection{Piezo element}
\begin{figure}[t]
\centering
\subfigure[View from above]{
\includegraphics[width=.45\columnwidth]{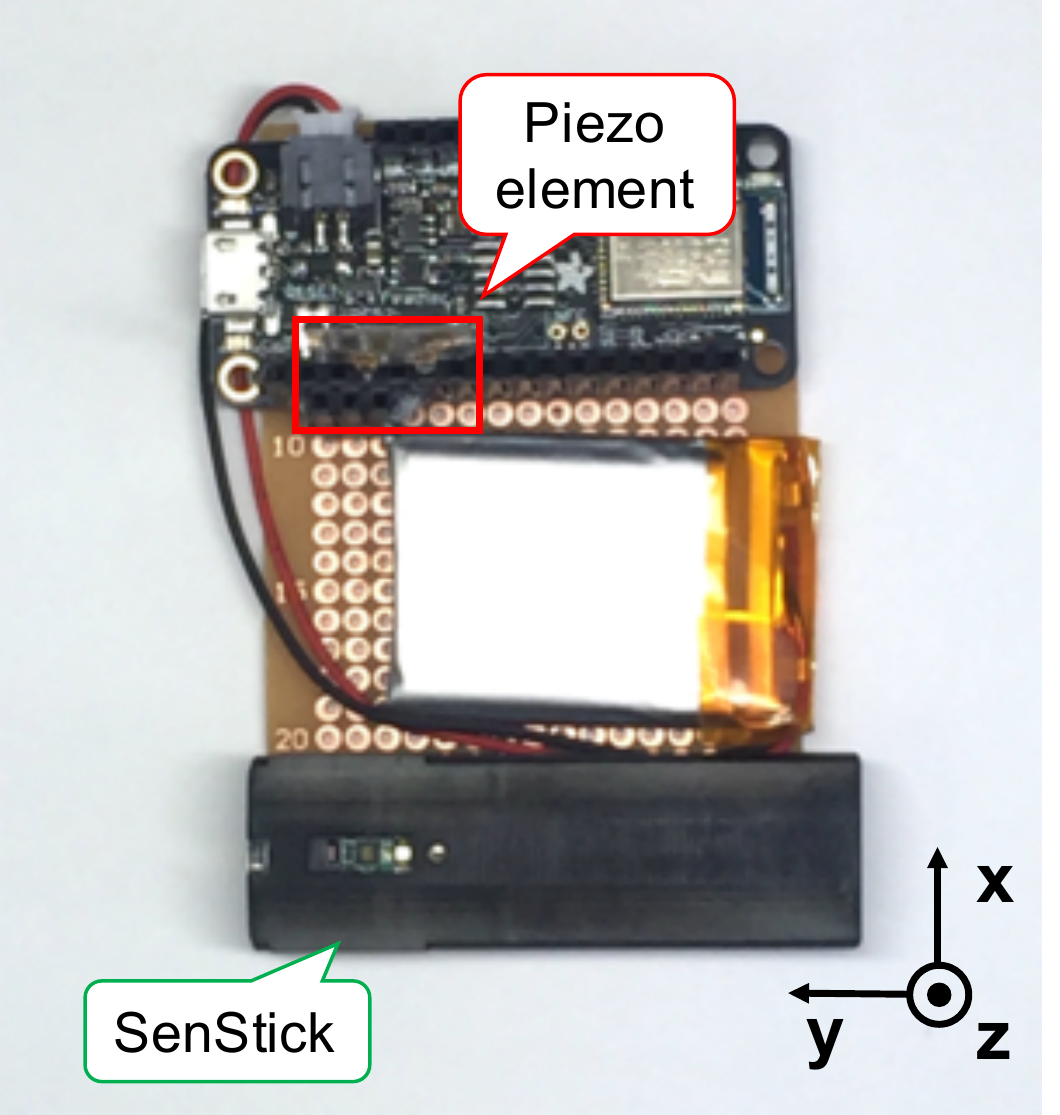}
}
\subfigure[View from upper diagonal]{
\includegraphics[width=.45\columnwidth]{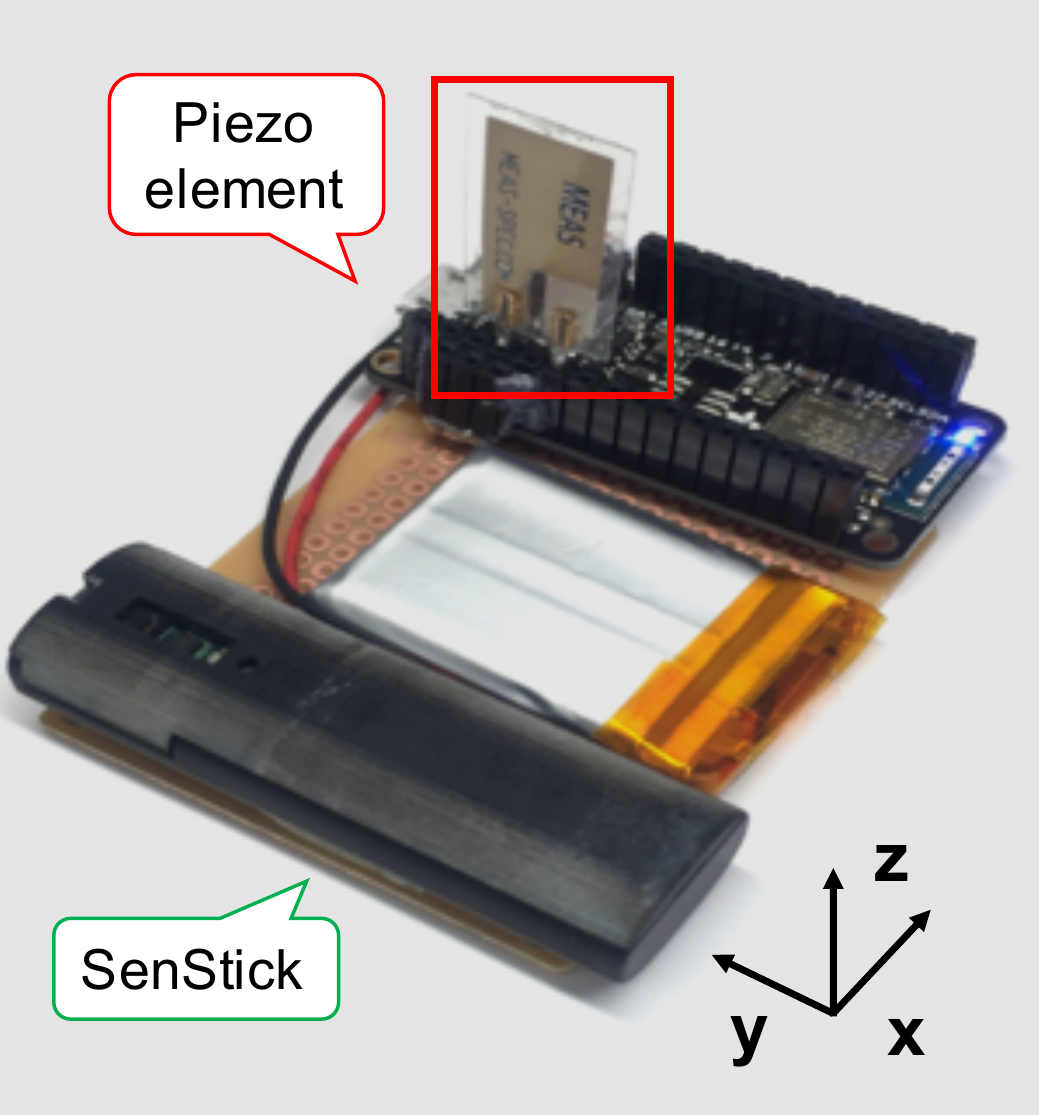}
}
\caption{A measurement system of piezo and 9-axis motion}
\label{Vib}
\end{figure}

To investigate the characteristics of piezo elements, we developed a measurement system (Fig. \ref{Vib}) equipped with a piezo element and SenStick, which can measure the generated-electricity of a piezo element and 9-axis motion simultaneously.

\begin{figure}[t]
\centering
\subfigure[Acceleration]{
\includegraphics[width=.45\columnwidth]{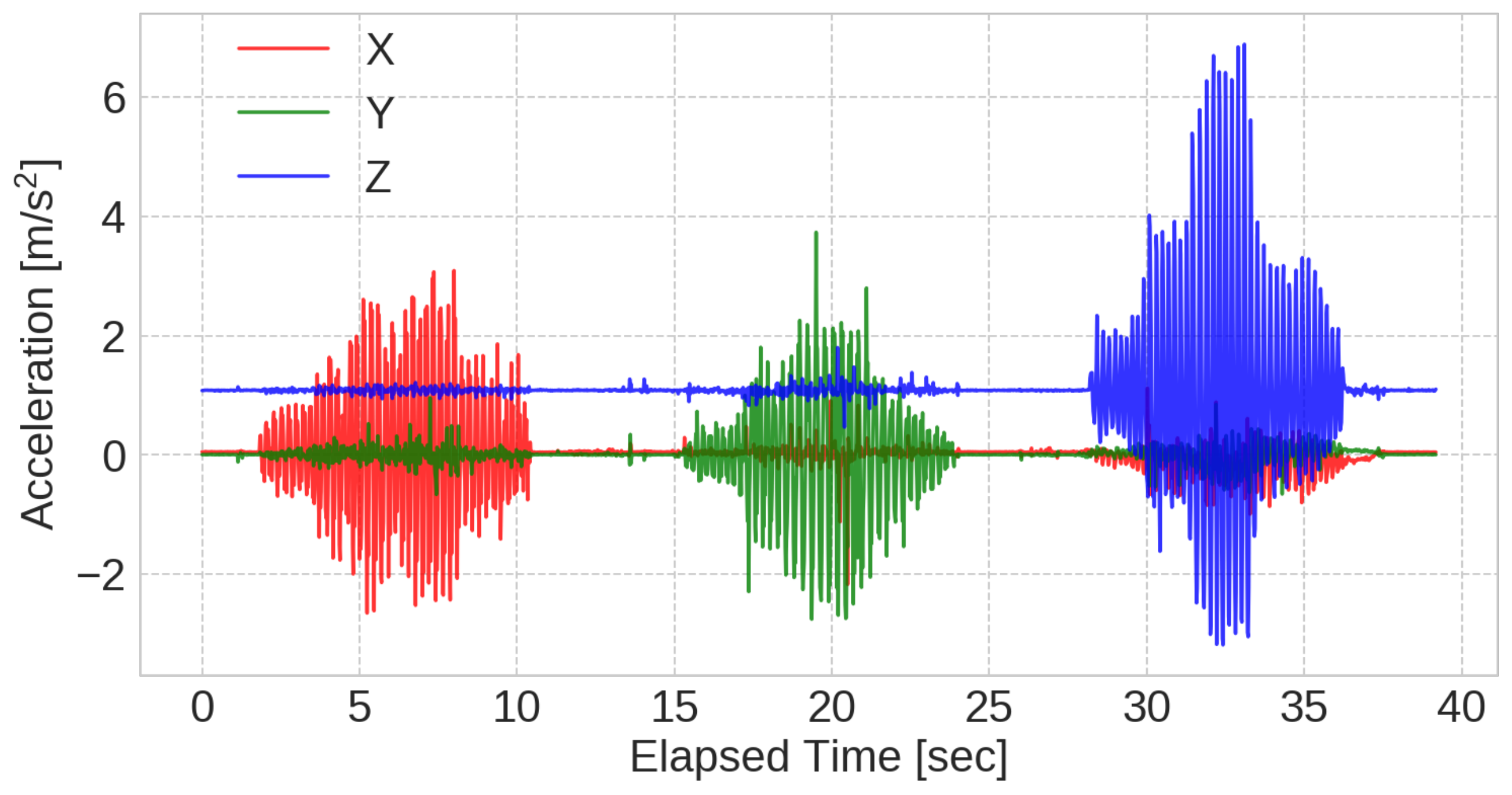}
\label{XYZ}
}
\subfigure[Generated electricity]{
\includegraphics[width=.45\columnwidth]{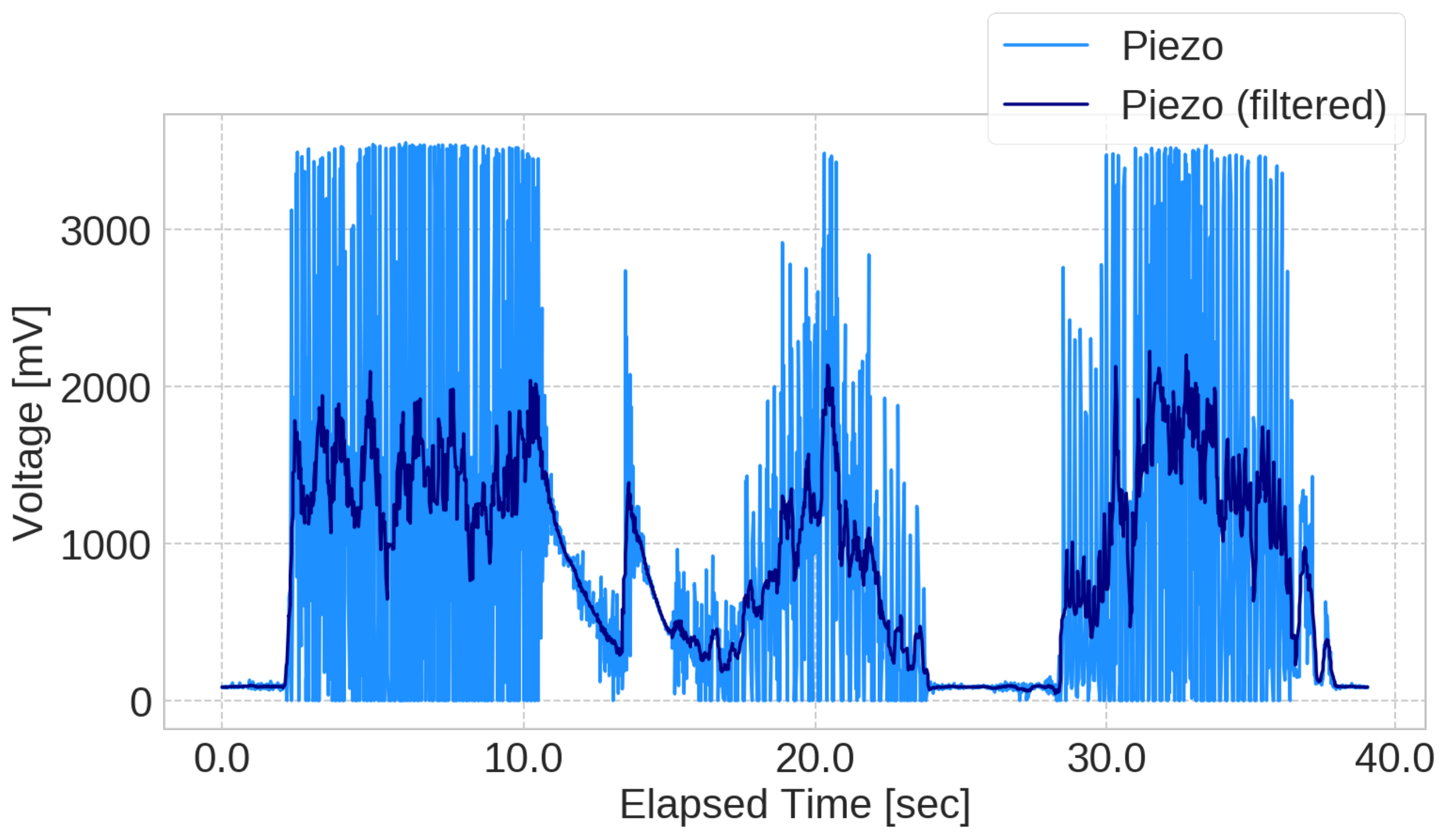}
\label{Piezo}
}
\caption{Characteristic of the piezo element against the motion}
\end{figure}

In this experiment, we installed LDT0-028K, a piezoelectric film of Measurement Specialties, Inc., parallel to the Y-Z plane in the measurement system. This means that the piezo element will curve in the X-axis direction according to the movement. Fig. \ref{XYZ} shows the result of acceleration values of the X, Y, Z axes. Fig. \ref{Piezo} shows the generated electricity of the piezo element, which includes raw data and the data smoothed by the moving average filter (time window is 0.2 sec).

This result suggests that the generated electricity of a piezo element has some correlation with the movement of the object and can be considered as a sensor for detecting motion. 

\subsection{Peltier element}
\begin{figure}[t]
\centering
\subfigure[Overall view]{
\includegraphics[width=.45\columnwidth]{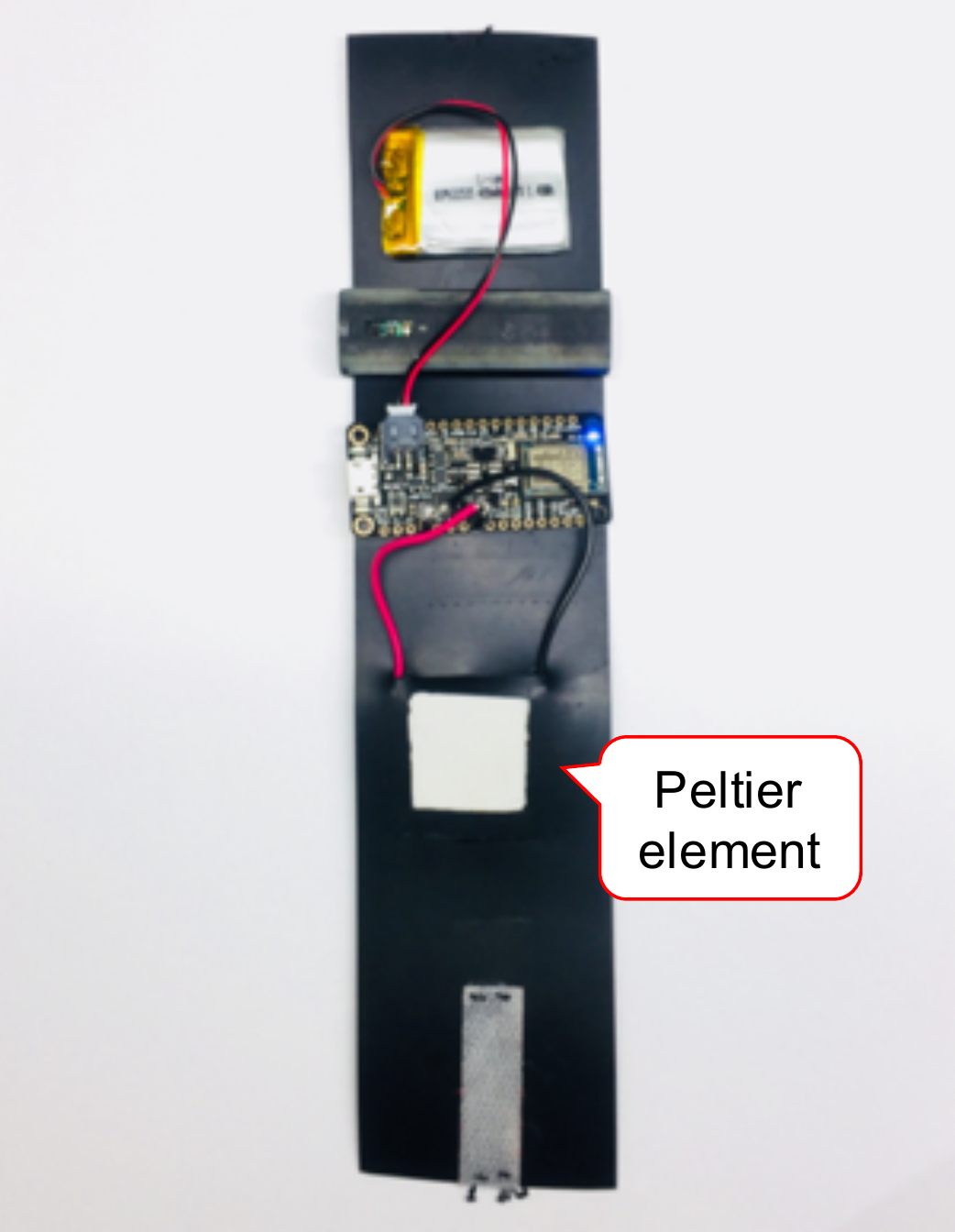}
}
\subfigure[Wearing view]{
\includegraphics[width=.45\columnwidth]{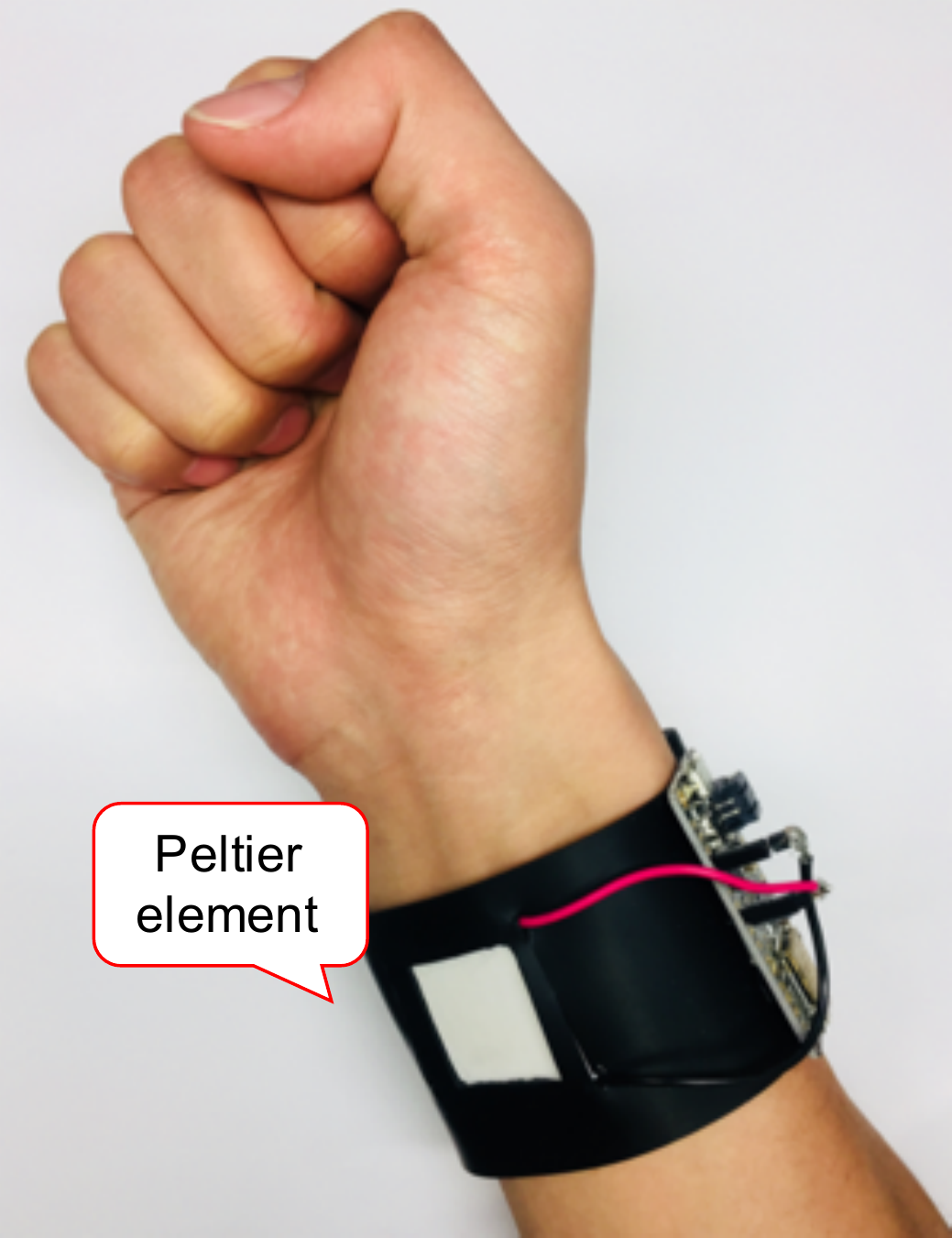}
}
\caption{A wearable measurement system for peltier element}
\label{Wrist}
\end{figure}

In the investigation of peltier elements, we observed the generated electricity by moving between indoors and outdoors. Fig. \ref {Wrist} shows the developed wearable system for this experiment, which equips a peltier element inside a band. The temperatures of the atmosphere and skin surface are measured by SenStick and Empatica’s E4 wristband\footnote{Real-time physiological signals | E4 EDA/GSR sensor :\\ \url{https://www.empatica.com/research/e4/}}, respectively.

\begin{figure}[t]
\centering
\subfigure[Temperature]{
\includegraphics[width=.45\columnwidth, angle=0]{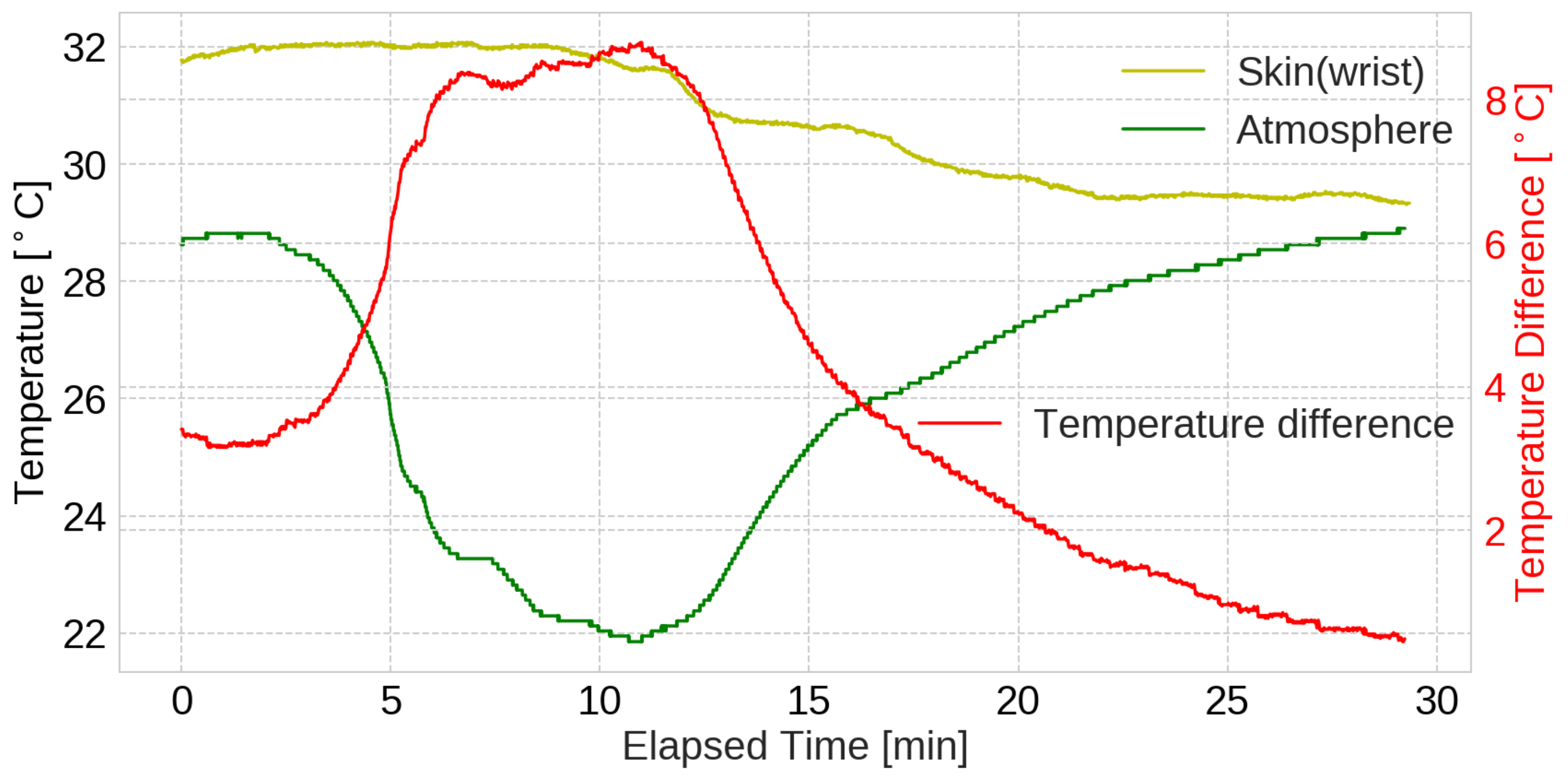}
\label{Temp}
}
\subfigure[Generated electricity]{
\includegraphics[width=.45\columnwidth, angle=0]{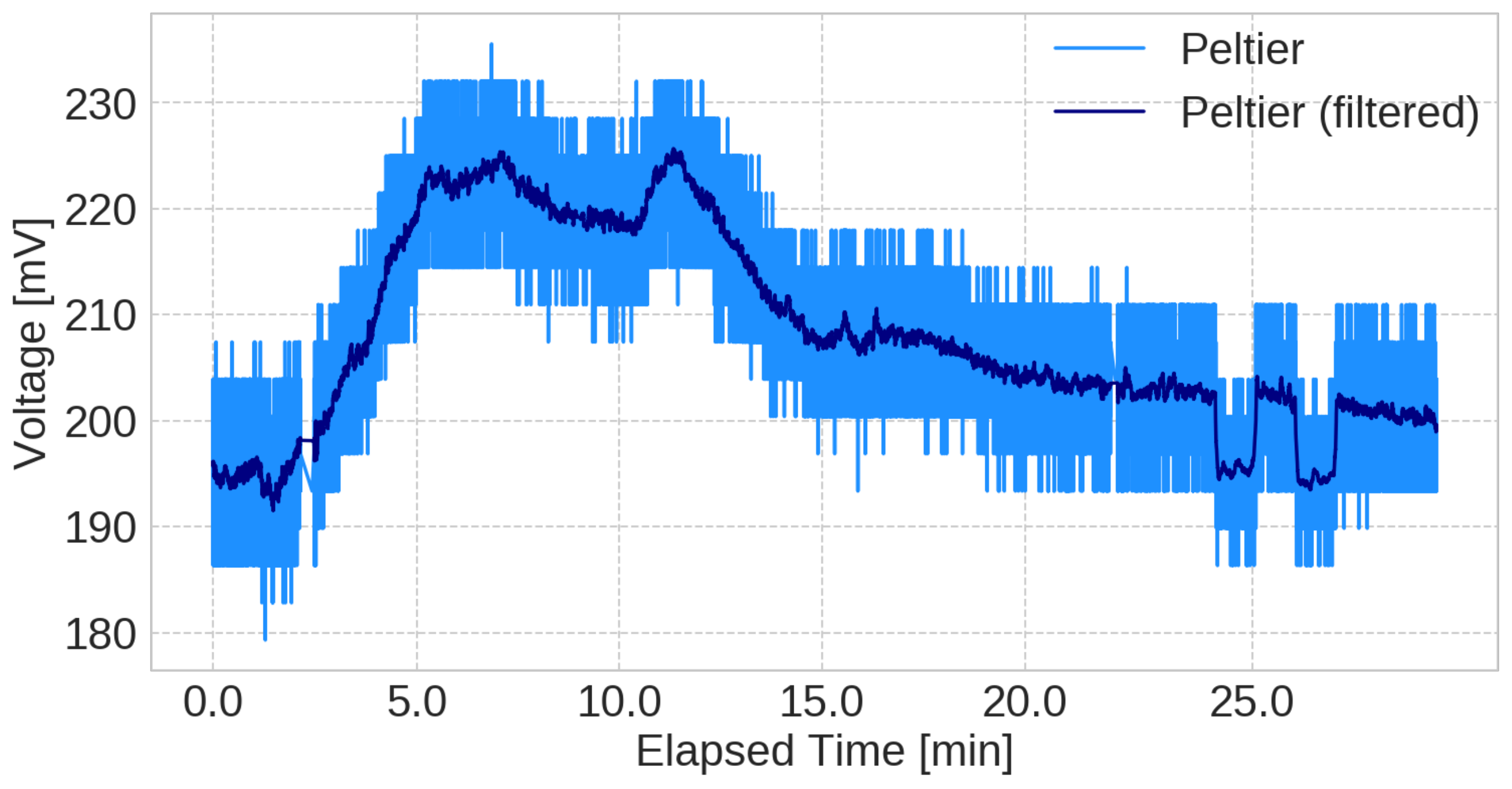}
\label{Peltier}
}
\caption{Characteristic of the peltier element against the temperature difference}
\end{figure}

Fig. \ref{Temp} represents the transition in skin temperature and ambient temperature. Fig. \ref{Peltier} shows the generated electricity of the peltier element, the raw data, and the data smoothed by the moving average filter (2 sec of time window). We can observe a high correlation between the temperature change and the generated electricity. This suggests that the peltier element can be used as a sensor to distinguish indoors or outdoors if we assume that the skin temperature is almost constant.

\subsection{Summary of measurement results}
We summarize the preliminary experiment as follows.

\begin{itemize}
\item Solar Cells\\
Since the characteristics against brightness and wavelength are different depending on the material, a combination of several types of SCs is expected to work well for place recognition. 

\item Piezo element\\
Since it works like accelerometer, a piezo element can be used to recognize the movement of a human. However, it may not be suitable for places where the user stays stationary. 

\item Peltier element\\
A peltier element can be used for indoor or outdoor detection. However, the peltier element has a usage restriction in that it must be in contact with the skin.
\end{itemize}

In this section, we investigated the capability of each EH element. In the following sections, we try to develop a place recognition system by building a sensor device and evaluating it through experiments. In particular, considering the use cases in offices or workplaces, we created a nameplate-like sensor device. Following the knowledge obtained in this section, we loaded only multiple solar cells and a piezo element on the device\footnote{\textcolor{black}{The peltier element was excluded from the device because it has restrictions on hardware implementation with respect to the available environment and shape of devices. First, it does not work in a hot environment such as in summer, because it always needs a difference between the temperatures of the skin and the air to generate electricity. Furthermore, it also has to keep contact with the skin and suitable shapes of devices for continuous operation is limited for office use. These are unsuitable properties for a nameplate-like device.}}. Furthermore, we implemented all the EH elements except for the peltier element to find the best combination of EH elements.

\section{A place recognition system with energy harvesters}
\label{place recognition}
Based on the result of the preliminary measurements, we design a place recognition system where several types of energy harvesters are used as a sensor. We call it EHAAS, hereafter. 

\subsection{Assumed lifelogging service} 
The goal of EHAAS is to realize long-term lifelogging composed of a time series of visited places in the working environment. The right side of Fig. \ref{system_ach} shows an image of the output of EHAAS. 

\subsection{EHAAS architecture} 
\begin{figure}[t]
\centering
\includegraphics[width=\columnwidth]{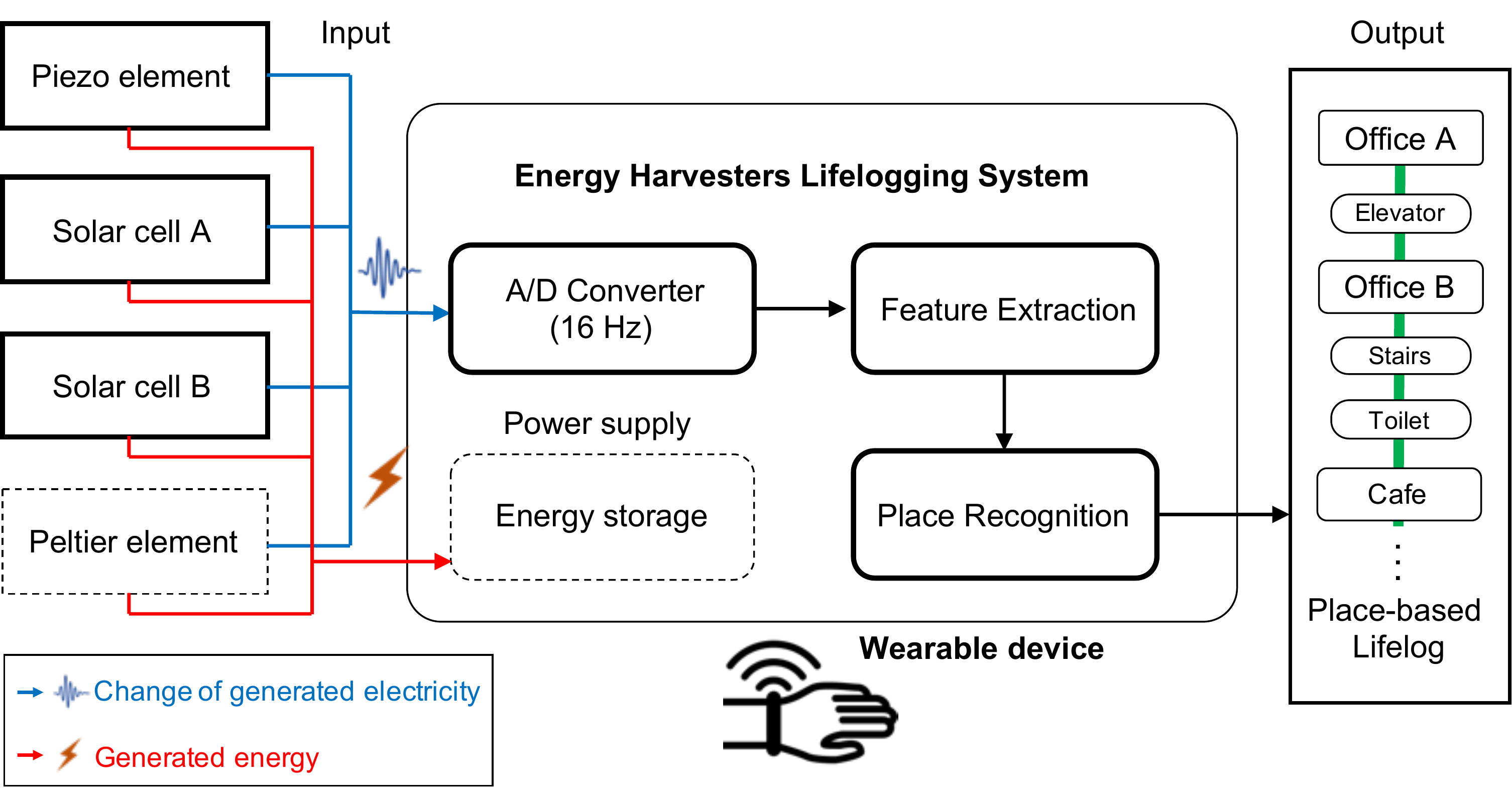}
\caption{Proposed EHAAS architecture}
\label{system_ach}
\end{figure}

EHAAS consists of a hardware component and a software component (algorithm). The former utilizes Bluefruit of Adafruit as the main board, to which multiple energy harvesting elements are connected as shown in the left side of Fig. \ref{system_ach}. The electricity generated by the energy harvesting elements is used for both energy and the sensor. 

\begin{figure*}[h]
\centering
\vspace{-3mm}
\includegraphics[width=2 \columnwidth]{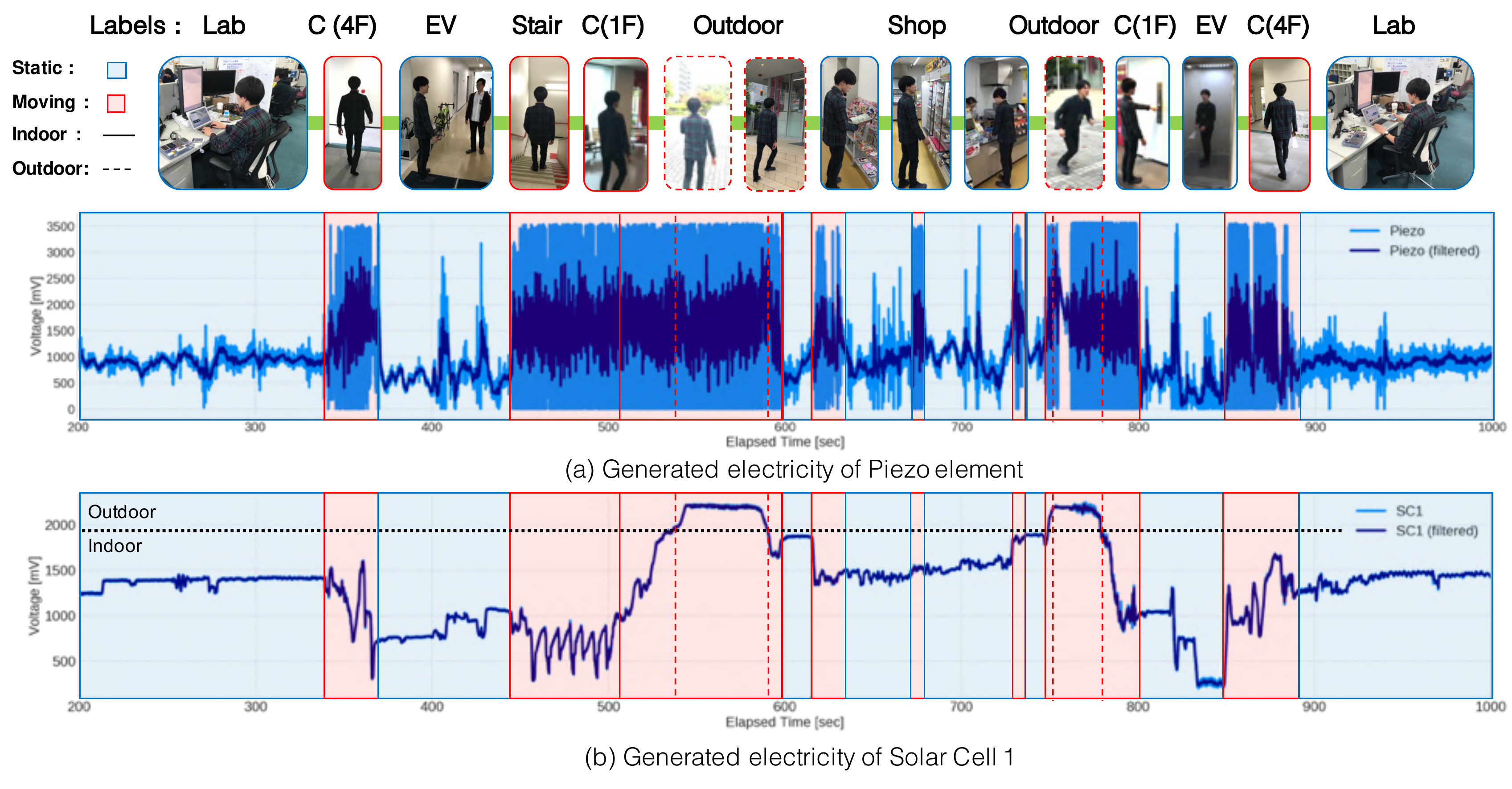}
\caption{Example of collected data}
\vspace{-3mm}
\label{timeseries}
\end{figure*}

In the main board, the following four modules are implemented. 

\subsubsection{A/D converter} 
All the generated electricity is input into an A/D converter on the board. The resolution and sampling rate of the A/D converter on Adafruit Bluefruit are 10-bit (0…1023) and 16Hz respectively. The input voltage is adjusted to below 3.6V by the proper resistance. This means that every digit returned from the ADC is 3.515625mV (= 3600mV/1024).

\subsubsection{Feature extraction} 
\label{B: Feature extraction}
We use several types of solar cells and a piezo element for extracting the features of places because the amount of generated electricity will depend on the environment such as light intensity, light wavelength, temperature, and the wearer’s movement. This module generates feature vectors composed of the generated electricity of each element every 2 seconds. 

\subsubsection{Place recognition} 
This module classifies the places based on a movement feature and an environmental feature. Here, the movement feature is abstracted into the same length as the environmental feature and combined. Finally, the classifier generated by machine learning decides the most probable place. The details will be explained in the next subsection. 

\subsubsection{Energy storage} 
All the generated electricity is charged into energy storage, such as a capacitor. However, the implementation of this part is beyond the scope of this paper. 

\subsection{Place recognition model}
We construct the model for place recognition by using machine learning. Through preliminary investigations, we employ Random Forests (RF) \cite{breiman2001random} as the best among various machine learning algorithms. 

Training data are needed for constructing the model. The feature used as the training data is the amount of generated electricity. This is obtained as the movement feature or the environmental features. As features, statistical values: average, variance, sum, median, maximum, minimum, difference between maximum and minimum, are calculated for every 2 second interval. Also, place labels are assigned to each interval as ground truth. 

It is difficult to construct a model that can adapt to all possible places, because people stay/pass various places in daily life. However, one of the goals of this system is lifelogging in a building (e.g., workplace, school, etc.), and the number of places where a user often stays in such an indoor environment is limited. We assume the number of places visited by one user is no more than 10, such as an office, meeting room, toilet, stair, corridor, elevator, outdoor, cafeteria, and so on. Therefore, a training dataset could be prepared in each environment, and in this paper it was collected at representative places in our university. 

Additionally, the training data should be collected considering various situations such as different time periods (morning, evening, night) and for different weather (sunny, cloudy, rainy, etc). Because our target is to recognize a place, not the behavior of each person, we do not need to develop each person’s model. Based on this, we believe that it is sufficient to collect 9 datasets that consider time and weather differences from no more than 10 places.

\section{Performance evaluation}
\label{performance evaluation}
We evaluate the usefulness and accuracy of our proposed system through experiment. 

\subsection{Evaluation environment and scenario}
As a target environment, we select our university campus where staff and students move among several buildings, rooms, and facilities within a stay. There are various kinds of places on the university campus, such as classrooms, laboratories, toilets, shops, and cafeterias. However, in such an environment, GPS cannot distinguish the room-level places. Even when walking outdoors between buildings, it is hard to receive GPS signals and calculate the position because the duration spent outdoors is too short. Fig. \ref{timeseries} shows a concrete movement scenario and the collected data. In this experiment, we target the following nine places: Laboratory, Seminar room, Toilet, Stairs, Elevator, Corridor (1st floor), Corridor (4th floor), Outdoors, and Cafeteria. In Fig.\ref{timeseries} (a) and (b), we show that each place generates a unique amount of electricity according to its environment or the movement of a user, and EH devices have a big potential for recognizing places.

\subsection{Data collection}
\begin{figure}[t]
\centering
\includegraphics[width=.6\columnwidth]{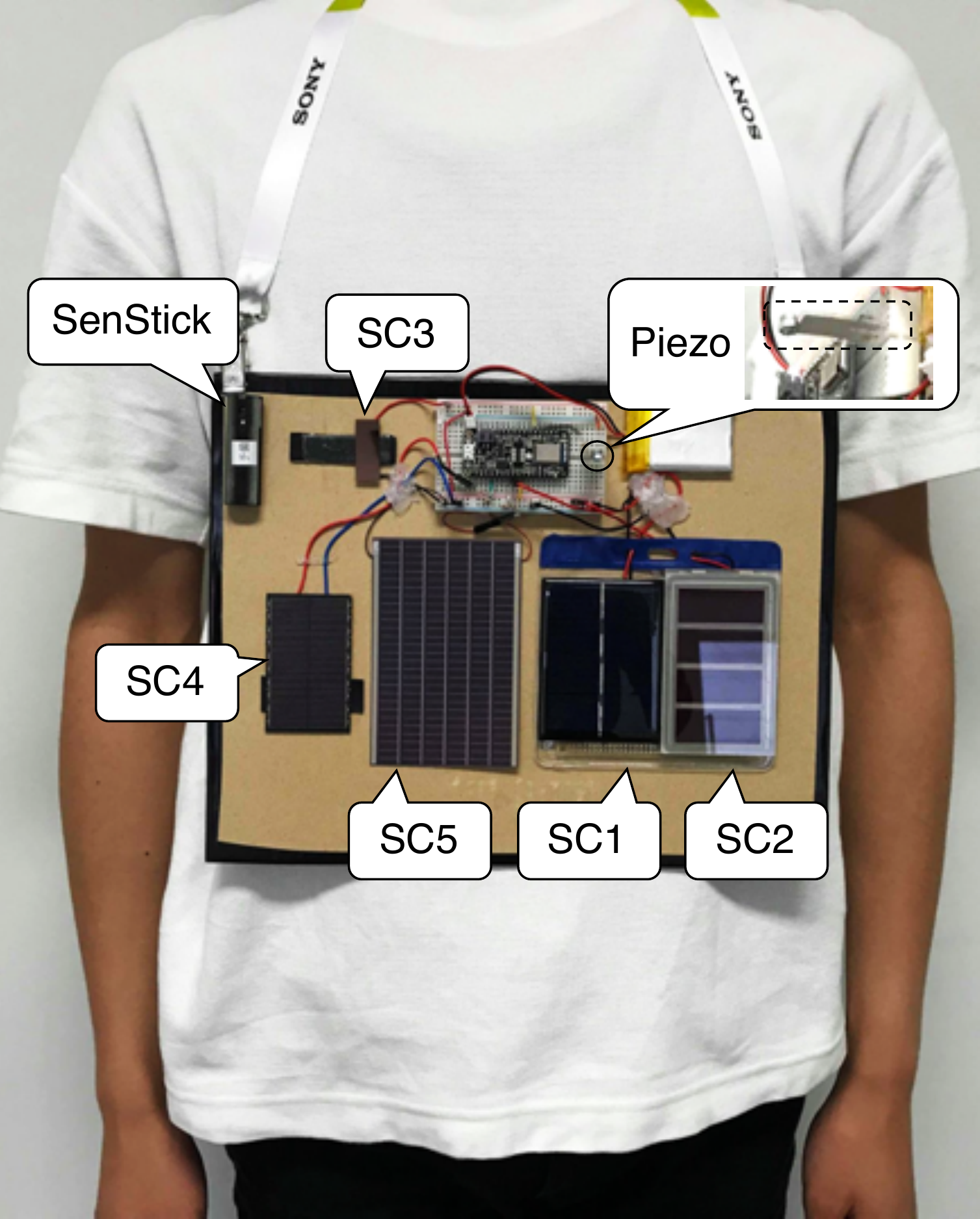}
\caption{Prototype wearable device}
\label{proto}
\end{figure}

To collect the data, we developed an experimental board where a microprocessor and several energy harvesting elements are affixed on the surface, as shown in Fig. \ref{proto}. Five different kinds of solar cells are attached to the board horizontally. A piezo element is set at the center. We adjust the weight attached to the tip of the piezo element to generate a moderate vibration (electricity) according to the body movement. We attached SenStick for measuring ground truth data of acceleration, gyro, magnetic, temperature, and light intensity.

We collected data in three days with different weather and at different times. The first day was rainy, the second day was cloudy, and the third day was sunny. The data was collected in different periods of time such as morning, afternoon, and evening including night. A data collector subject wearing this device visited nine places by walking along a predetermined route. Finally, we collected 10 sets of data in different situations. 

We calculated seven features (average, variance, sum, median, maximum, minimum, difference between maximum and minimum) every two seconds from the collected data. The window size (two seconds) was decided empirically. As a result, a dataset with 42 features (seven features $\times$ six elements) was created. Moreover, the ground truth label was applied by checking the video recorded in the experiment. Similarly to the features, we assigned place labels for every 2 seconds. We performed the same process for all the collected data and created ten case datasets.

\subsection{Evaluation method}
We used a leave-one-case-out cross-validation for the evaluation. Out of the ten cases collected in different situations, one case was excluded as a test dataset for evaluation, and the remaining datasets were used for constructing a place recognition model by machine learning (Random Forest). We carried out this process for each of the 10 datasets and evaluated the accuracy of our proposed place recognition system, trying to avoid overfitting and minimize generalization errors. Also, we evaluated the effectiveness of this method by comparing the accuracy with that derived by a conventional method based on an illuminance sensor and accelerometer. 

In addition, we compared the accuracy among different combinations of energy harvesters to investigate the possibility of reducing the number of elements. In particular, the combination of solar cells is important because it inhibits miniaturization. As we intend to apply EHAAS to a wearable system, a smaller number of elements will be better for implementation. 

\subsection{Result}
\begin{table}[t]
\centering
\caption{place recognition accuracy of the proposed system (sensor: All energy harvesting elements)}
\label{PRF_data1}
\scalebox{0.9}{
{\renewcommand{\arraystretch}{1.3}
\begin{tabular}{l|c|c|c|r}
\hline
              & Precision & Recall & F-value & Num. of data \\ \hline\hline
Lab          	& 	0.969	 & 	0.944	& 	0.957	& 	7899	\\ \hline
Seminar\_Room	 &	0.701	 &	0.752	 &	0.725	& 	1767	  \\ \hline
Cafeteria    	 &	0.837	 &	0.725	 &	0.777	 &	945	 \\ \hline
Corridor\_4th	 &	0.537	 &	0.711	 &	0.612	 &	464	 \\ \hline
Toilet       	 &	0.955	 &	0.823	 &	0.884	 &	434	 \\ \hline
Corridor\_1st	 &	0.635	 &	0.738	 &	0.683	 &	427	 \\ \hline
Stairs       	 &	0.764	 &	0.86	 &	0.809	 &	350	 \\ \hline
Outdoors      	 &	0.97	 &	0.855	 &	0.909	& 	345	 \\ \hline
EV           	 &	0.721	 &	0.913	 &	0.805	 &	161	 \\ \hline\hline
avg / total  	 &	0.887	 &	0.877	 &	0.88	 &	12792	 \\ \hline
\end{tabular}
}
}
\end{table}

\begin{table}[t]
\centering
\caption{\textcolor{black}{place recognition accuracy of the conventional system (sensor: illuminance)}}
\label{Senstick_ill}
\scalebox{0.9}{
{\renewcommand{\arraystretch}{1.3}
\begin{tabular}{l|c|c|c|r}
\hline
              	& 	Precision	& 	Racall	& 	F-value	& 	Num. of data	\\ \hline\hline
Lab          	& 	0.872	 & 	0.896	& 	0.883	& 	6867	\\ \hline
Seminar\_Room	 &	0.418	 &	0.350	 &	0.381	& 	1416	  \\ \hline
Cafeteria    	 &	0.402	 &	0.367	 &	0.384	 &	1340	 \\ \hline
Corridor\_4th	 &	0.450	 &	0.508	 &	0.478	 &	366	 \\ \hline
Toilet       	 &	0.483	 &	0.413	 &	0.445	 &	516	 \\ \hline
Corridor\_1st	 &	0.435	 &	0.477	 &	0.455	 &	384	 \\ \hline
Stairs       	 &	0.600	 &	0.661	 &	0.629	 &	313	 \\ \hline
Outdoors      	 &	0.623	 &	0.845	 &	0.718	& 	317	 \\ \hline
EV           	 &	0.410	 &	0.423	 &	0.416	 &	182	 \\ \hline\hline
avg / total  	 &	0.697	 &	0.707	 &	0.701	 &	11701	 \\ \hline
\end{tabular}
}
}
\end{table}


\begin{table}[t]
\centering
\caption{\textcolor{black}{place recognition accuracy of the conventional system (sensor: acceleration and illuminance)}}
\label{Senstick_acc_ill}
\scalebox{0.9}{
{\renewcommand{\arraystretch}{1.3}
\begin{tabular}{l|c|c|c|r}
\hline
              	& 	Precision	& 	Racall	& 	F-value	& 	Num. of data	\\ \hline\hline
Lab          	& 	0.911	 & 	0.869	& 	0.889	& 	7452	\\ \hline
Seminar\_Room	 &	0.63	 &	0.656	 &	0.643	& 	1534	  \\ \hline
Cafeteria    	 &	0.451	 &	0.531	 &	0.488	 &	1516	 \\ \hline
Corridor\_4th	 &	0.584	 &	0.602	 &	0.593	 &	405	 \\ \hline
Toilet       	 &	0.632	 &	0.514	 &	0.567	 &	560	 \\ \hline
Corridor\_1st	 &	0.55	 &	0.648	 &	0.595	 &	421	 \\ \hline
Stairs       	 &	0.834	 &	0.856	 &	0.845	 &	353	 \\ \hline
Outdoors      	 &	0.88	 &	0.872	 &	0.876	& 	344	 \\ \hline
EV           	 &	0.597	 &	0.667	 &	0.63	 &	189	 \\ \hline\hline
avg / total  	 &	0.781	 &	0.769	 &	0.774	 &	12774	 \\ \hline
\end{tabular}
}
}
\end{table}


\begin{table}[t]
\centering
\caption{Average F-value of each combination in the proposed system}
\label{result_overview}
{\renewcommand{\arraystretch}{1.3}
\scalebox{0.9}{
\begin{tabular}{c|c|c|c|c|c}
\hline
\multicolumn{2}{c|}{ Num. of EHs } & Combination & Precision & Recall & F-value \\\hline\hline
\multicolumn{2}{c|}{ All (6 types) } & 1,2,3,4,5,p & 0.887 & 0.877 & 0.880 \\\hline
\multirow{2}{*}{ 5 types } & Best & 2,3,4,5,p & 0.89 & 0.883 & \textbf{0.885} \\\cline{2-6}
& Worst & 1,2,3,4,p & 0.883 & 0.867 & 0.872 \\\hline
\multirow{2}{*}{ 4 types } & Best & 1,2,3,5 & 0.89 & 0.877 & 0.882 \\\cline{2-6}
& Worst & 1,4,5,p & 0.843 & 0.835 & 0.836 \\\hline
\multirow{2}{*}{ 3 types } & Best & 3,4,5 & 0.883 & 0.875 & 0.878 \\\cline{2-6}
& Worst & 2,4,p & 0.785 & 0.785 & 0.784 \\\hline
\multirow{2}{*}{ 2 types } & Best & 1,2 & 0.874 & 0.855 & \textbf{0.862} \\\cline{2-6}
& Worst & 2,p & 0.745 & 0.763 & 0.751 \\\hline
\multirow{2}{*}{ 1 type } & Best & 3 & 0.763 & 0.768 & 0.762 \\\cline{2-6}
& Worst & p & 0.644 & 0.686 & 0.655 \\\hline
\end{tabular}
}
}
\vspace{0.5mm}
\begin{flushleft}
\footnotesize
*p: piezo element, Numerical number : type of solar cell. 1 means SC1. 
\end{flushleft}
\end{table}

First, in Table \ref{PRF_data1}, we show the result in the case that all the EH elements are used as a sensor. We see that our system could accurately recognize Lab with 95.9\% of F-value and Outdoors with a 91.0\% F-value. However, the F-value of both corridors on the 1st floor and 4th floor are a bit lower (71.8\% and 61.4\%) than other places because the light condition in corridors in the same building is similar. The average F-value for all of the nine places is 88.3\%, which we believe is high enough for practical use, compared with the accuracy (90.2\% in average) achieved by BLE-based room-level localization \cite{kyritsis2016ble}. We also evaluated the place recognition accuracy of a conventional system\cite{Uwe2006} where an illuminance sensor is used as a sensor, in combination with an acceleration sensor for the recognition. Table \ref{Senstick_ill} shows the result for the conventional system that utilizes an illuminance sensor (Photocell/Photodiode) for place recognition. It was able to recognize the place ``Lab'' with 88.3\% accuracy. However, the accuracy of other places is low. Overall, the average F-value for all of the nine places is 70.1\%. Table \ref{Senstick_acc_ill} shows the result of the system where both an acceleration sensor and an illuminance sensor are used for place recognition. The average F-value for all of the nine places is 77.4\%. Except for ``Stairs'', where the place recognition accuracy is lower than our proposed system. We think that the information of the acceleration sensor worked effectively for detecting Stairs. However, it did not work well for recognizing other places.

Next, we compare the F-value for different combinations of EH elements to investigate the deterioration of the F-value when reducing the number of EH elements used. We investigated the performance of all the combinations of six EH elements, 63 patterns in total. Table \ref{result_overview} shows the results, where we pick up only the best result and the worst result for each number of EH elements used. 

We see that the F-value is basically reduced when the number of elements is reduced. The best performance, 88.5\%, was achieved by the combination of five EH elements (SC2, SC3, SC4, SC5, piezo). On the other hand, the worst performance, 65.5\%, was achieved in the case where we used one piezo element. However, if we add one more EH element, the F-value rises to 86.2\%, where SC1 and SC2 are used. The performance difference against the best combination (which requires 6 elements) is just 2.3\%. This result is very convincing because we observed that SC1 (Polycrystalline silicon) and SC2 (Organic thin film) have different characteristics against the illuminance and wavelength in the preliminary experiment (see Fig. \ref{intensity-result-a} and Fig. \ref{QE}). From this result, we can say that the combination of a couple of EH elements having different characteristics have a high capability to be used as a sensor for place recognition. Furthermore, this result indicates the possibility of the miniaturization of devices for practical use of this system.

Table \ref{PRF_data9} shows the detailed result of place recognition in the case where we used only SC1 and SC2 as a sensor. Compared with Table \ref{PRF_data1}, we see that the recognition accuracy of Corridor\_1st and Corridor\_4th are degraded. These places are also hard to distinguish for other combination of EH elements because of the similarity of the environment (light condition). However, the place recognition accuracy for other places are almost the same despite the number of EH elements being reduced from six to two. Hence, we consider that the most reasonable combination for wearables is SC1 and SC2. 

\begin{table}[t]
\centering
\caption{\textcolor{black}{place recognition accuracy of the proposed system (sensor: SC1 and SC2)}}
\label{PRF_data9}
\scalebox{0.9}{
{\renewcommand{\arraystretch}{1.3}
\begin{tabular}{l|c|c|c|r}
\hline
              & Precision & Recall & F-value & Num. of data \\ \hline\hline
Lab & 0.97 & 0.92 & 0.944 & 7899 \\\hline
Seminar\_Room & 0.635 & 0.761 & 0.692 & 1767 \\\hline
Toilet & 0.944 & 0.783 & 0.856 & 434 \\\hline
Stairs & 0.685 & 0.814 & 0.744 & 350 \\\hline
EV & 0.649 & 0.814 & 0.722 & 161 \\\hline
Corridor\_1st & 0.638 & 0.677 & 0.657 & 427 \\\hline
Corridor\_4th & 0.452 & 0.651 & 0.534 & 464 \\\hline
Outdoors & 0.925 & 0.861 & 0.892 & 345 \\\hline
Cafeteria & 0.894 & 0.726 & 0.801 & 945 \\\hline\hline
avg / total & 0.874 & 0.855 & 0.862 & 12792 \\\hline
\end{tabular}
}
}
\end{table}

Finally, we discuss the effect of the piezo element on the place recognition performance. The result, 65.5\% of average F-value, was not good in this experiment. We expected that the piezo element would help to improve the performance when it was used with other EH elements. Although we did not assume a no-light environment in this experiment, using a piezo element as a sensor will be useful for environments with no light where the solar cell does not generate electricity at all.

\section{Conclusion}
\label{conclusion}
In this paper, we have proposed a novel place recognition system for wearables called EHAAS (Energy Harvesters As A Sensor), where multiple energy harvesting elements are used as a sensor. First, we investigated various EH elements to clarify the characteristics of those elements against the environment. Second, we implemented a prototype wearable device and proposed a place recognition algorithm. Finally, we showed that our proposed EHAAS achieved 88.5\% accuracy with five EH elements and 86.2\% with just two EH elements. Furthermore, we confirmed that this system has better accuracy than an existing system consisting of an illuminance sensor and accelerometer, in place recognition in the daily life of indoor workers. This means that we can implement an EHAAS-based system on a smaller wearable such as a watch (Fig. \ref{Wrist}). 

In this paper, we did not use EH element as an energy source. As a next step, we will design a circuit that uses EH elements as both an energy source and a sensor. Also, we will investigate other combinations of EH elements including peltier elements and rectenna\cite{georgiadis2010rectenna}.

\section*{Acknowledgment}
This work was supported by JSPS Grants-in-Aid for Scientific Research (KAKENHI) Grant Number 18H03233.

\bibliographystyle{SIGCHI-Reference-Format}
\bibliography{sample-bibliography-sigchi-a}


\begin{thebibliography}{00}


\ifx \showCODEN    \undefined \def \showCODEN     #1{\unskip}     \fi
\ifx \showDOI      \undefined \def \showDOI       #1{{\tt DOI:}\penalty0{#1}\ }
  \fi
\ifx \showISBNx    \undefined \def \showISBNx     #1{\unskip}     \fi
\ifx \showISBNxiii \undefined \def \showISBNxiii  #1{\unskip}     \fi
\ifx \showISSN     \undefined \def \showISSN      #1{\unskip}     \fi
\ifx \showLCCN     \undefined \def \showLCCN      #1{\unskip}     \fi
\ifx \shownote     \undefined \def \shownote      #1{#1}          \fi
\ifx \showarticletitle \undefined \def \showarticletitle #1{#1}   \fi
\ifx \showURL      \undefined \def \showURL       #1{#1}          \fi

\bibitem{bharatula2004towards}
{Nagendra~B Bharatula}, {Stijn Ossevoort}, {Mathias St{\"a}ger}, {and} {Gerhard
  Tr{\"o}ster}. 2004.
\newblock \showarticletitle{Towards wearable autonomous microsystems}. In {\em
  International Conference on Pervasive Computing}. Springer, 225--237.
\newblock


\bibitem{4812905}
{M. Bouet} {and} {A.~L. dos Santos}. 2008.
\newblock \showarticletitle{RFID tags: Positioning principles and localization
  techniques}. In {\em 2008 1st IFIP Wireless Days}. 1--5.
\newblock
\showISSN{2156-9711}
\showDOI{%
\url{http://dx.doi.org/10.1109/WD.2008.4812905}}


\bibitem{breiman2001random}
{Leo Breiman}. 2001.
\newblock \showarticletitle{Random forests}.
\newblock {\em Machine learning\/} {45}, 1 (2001), 5--32.
\newblock


\bibitem{brunato2005statistical}
{Mauro Brunato} {and} {Roberto Battiti}. 2005.
\newblock \showarticletitle{Statistical learning theory for location
  fingerprinting in wireless LANs}.
\newblock {\em Computer Networks\/} {47}, 6 (2005), 825--845.
\newblock


\bibitem{ciurana2007wlan}
{Marc Ciurana}, {Sebastiano Cugno}, {and} {Francisco Barcelo-Arroyo}. 2007.
\newblock \showarticletitle{WLAN indoor positioning based on TOA with two
  reference points}. In {\em Positioning, Navigation and Communication, 2007.
  WPNC'07. 4th Workshop on}. IEEE, 23--28.
\newblock


\bibitem{cypriani2009open}
{Matteo Cypriani}, {Fr{\'e}d{\'e}ric Lassabe}, {Philippe Canalda}, {and}
  {Fran{\c{c}}ois Spies}. 2009.
\newblock \showarticletitle{Open wireless positioning system: a Wi-Fi-based
  indoor positioning system}. In {\em Vehicular Technology Conference Fall (VTC
  2009-Fall), 2009 IEEE 70th}. IEEE, 1--5.
\newblock


\bibitem{akpa2018perhealth-backup}
{A.~H.~Akpa Elder}, {Fujiwara Masashi}, {Arakawa Yutaka}, {Suwa Hirohiko},
  {and} {Yasumoto Keiichi}. 2018.
\newblock \showarticletitle{GIFT: Glove for Indoor Fitness Tracking System}. In
  {\em Third IEEE PerCom Workshop on Pervasive Health
  Technologies(PerHealth2018)}. IEEE, 1--6.
\newblock


\bibitem{6042868}
{C. Feng}, {W.~S.~A. Au}, {S. Valaee}, {and} {Z. Tan}. 2012.
\newblock \showarticletitle{Received-Signal-Strength-Based Indoor Positioning
  Using Compressive Sensing}.
\newblock {\em IEEE Transactions on Mobile Computing\/} {11}, 12 (Dec 2012),
  1983--1993.
\newblock
\showISSN{1536-1233}
\showDOI{%
\url{http://dx.doi.org/10.1109/TMC.2011.216}}


\bibitem{georgiadis2010rectenna}
{A Georgiadis}, {G~Vera Andia}, {and} {A Collado}. 2010.
\newblock \showarticletitle{Rectenna design and optimization using reciprocity
  theory and harmonic balance analysis for electromagnetic (EM) energy
  harvesting}.
\newblock {\em IEEE Antennas and Wireless Propagation Letters\/}  {9} (2010),
  444--446.
\newblock


\bibitem{emotion}
{Knez Igor} {and} {Kers Christina}. 2000.
\newblock \showarticletitle{Effects of Indoor Lighting, Gender, and Age on Mood
  and Cognitive Performance}.
\newblock {\em Environment and Behavior\/} {32}, 6 (2000), 817--831.
\newblock


\bibitem{Kalantarian}
{H. Kalantarian} {and} {M. Sarrafzadeh}. 2016.
\newblock \showarticletitle{Pedometers Without Batteries: An Energy Harvesting
  Shoe}.
\newblock {\em IEEE Sensors Journal\/} {16}, 23 (Dec 2016), 8314--8321.
\newblock
\showISSN{1530-437X}
\showDOI{%
\url{http://dx.doi.org/10.1109/JSEN.2016.2591331}}


\bibitem{fatigue}
{C.H.J.Smolders Karin} {and} {Kort Yvonne, A.W.de}. 2014.
\newblock \showarticletitle{Bright light and mental fatigue: Effects on
  alertness, vitality, performance and physiological arousal}.
\newblock {\em Journal of Environmental Psychology\/}  {39} (2014), 77--91.
\newblock


\bibitem{khalifa2017harke}
{Sara Khalifa}, {Guohao Lan}, {Mahbub Hassan}, {Aruna Seneviratne}, {and}
  {Sajal~K Das}. 2017.
\newblock \showarticletitle{HARKE: Human Activity Recognition from Kinetic
  Energy Harvesting Data in Wearable Devices}.
\newblock {\em IEEE Transactions on Mobile Computing\/} (2017).
\newblock


\bibitem{kourogi2006indoor}
{Masakatsu Kourogi}, {Nobuchika Sakata}, {Takashi Okuma}, {and} {Takeshi
  Kurata}. 2006.
\newblock \showarticletitle{Indoor/outdoor pedestrian navigation with an
  embedded GPS/RFID/self-contained sensor system}.
\newblock In {\em Advances in Artificial Reality and Tele-Existence}. Springer,
  1310--1321.
\newblock


\bibitem{kyritsis2016ble}
{Athanasios~I Kyritsis}, {Panagiotis Kostopoulos}, {Michel Deriaz}, {and}
  {Dimitri Konstantas}. 2016.
\newblock \showarticletitle{A ble-based probabilistic room-level localization
  method}. In {\em Localization and GNSS (ICL-GNSS), 2016 International
  Conference on}. IEEE, 1--6.
\newblock


\bibitem{purpose}
{Bellia L.}, {Bisegna F.}, {and} {Spada G.} 2011.
\newblock \showarticletitle{Lighting in indoor environments: Visual and
  non-visual effects of light sources with different spectral power
  distributions}.
\newblock {\em Building and Environment\/} {46}, 10 (2011), 1984--1992.
\newblock


\bibitem{lee2015non}
{Seungwoo Lee}, {Yungeun Kim}, {Daye Ahn}, {Rhan Ha}, {Kyoungwoo Lee}, {and}
  {Hojung Cha}. 2015.
\newblock \showarticletitle{Non-obstructive room-level locating system in home
  environments using activity fingerprints from smartwatch}. In {\em
  Proceedings of the 2015 ACM International Joint Conference on Pervasive and
  Ubiquitous Computing}. ACM, 939--950.
\newblock


\bibitem{6129175}
{Hung-Huan Liu} {and} {Yu-Non Yang}. 2011.
\newblock \showarticletitle{WiFi-based indoor positioning for multi-floor
  Environment}. In {\em TENCON 2011 - 2011 IEEE Region 10 Conference}.
  597--601.
\newblock
\showISSN{2159-3450}
\showDOI{%
\url{http://dx.doi.org/10.1109/TENCON.2011.6129175}}


\bibitem{mautz2012indoor}
{Rainer Mautz}. 2012.
\newblock \showarticletitle{Indoor positioning technologies}.
\newblock  (2012).
\newblock


\bibitem{Nakamura:2017hk-backup}
{Yugo Nakamura}, {Yutaka Arakawa}, {Takuya Kanehira}, {Masashi Fujiwara}, {and}
  {Keiichi Yasumoto}. 2017.
\newblock \showarticletitle{{SenStick: Comprehensive Sensing Platform with an
  Ultratiny All-In-One Sensor Board for IoT Research}}.
\newblock {\em Journal of Sensors\/}  {2017} (2017), 16.
\newblock


\bibitem{otoda2018census-backup}
{Yasuhiro Otoda}, {Teruhiro Mizumoto}, {Yutaka Arakawa}, {Chihiro Nakajima},
  {Mitsuhiro Kohana}, {Motohiro Uenishi}, {and} {Keiichi Yasumoto}. 2018.
\newblock \showarticletitle{Census: Continuous posture sensing chair for office
  workers}. In {\em Consumer Electronics (ICCE), 2018 IEEE International
  Conference on}. IEEE, 1--2.
\newblock


\bibitem{pai2012padati}
{Deepak Pai}, {Inguva Sasi}, {Phani~Shekhar Mantripragada}, {Mudit Malpani},
  {and} {Nitin Aggarwal}. 2012.
\newblock \showarticletitle{Padati: A robust pedestrian dead reckoning system
  on smartphones}. In {\em Trust, Security and Privacy in Computing and
  Communications (TrustCom), 2012 IEEE 11th International Conference on}. IEEE,
  2000--2007.
\newblock


\bibitem{paradiso2005energy}
{Joseph~A Paradiso} {and} {Thad Starner}. 2005.
\newblock \showarticletitle{Energy scavenging for mobile and wireless
  electronics}.
\newblock {\em IEEE Pervasive computing\/} 1 (2005), 18--27.
\newblock


\bibitem{tarzia2011indoor}
{Stephen~P Tarzia}, {Peter~A Dinda}, {Robert~P Dick}, {and} {Gokhan Memik}.
  2011.
\newblock \showarticletitle{Indoor localization without infrastructure using
  the acoustic background spectrum}. In {\em Proceedings of the 9th
  international conference on Mobile systems, applications, and services}. ACM,
  155--168.
\newblock


\bibitem{Uwe2006}
{Maurer Uwe}, {Rowe Anthony}, {Smailagic Asim}, {and} {Siewiorek Daniel}. 2006.
\newblock \showarticletitle{Location and Activity Recognition Using eWatch: A
  Wearable Sensor Platform}.
\newblock {\em Ambient Intelligence in Everyday Life\/}  {3864} (2006),
  86--102.
\newblock


\bibitem{WOO20113}
{Sunkyu Woo}, {Seongsu Jeong}, {Esmond Mok}, {Linyuan Xia}, {Changsu Choi},
  {Muwook Pyeon}, {and} {Joon Heo}. 2011.
\newblock \showarticletitle{Application of WiFi-based indoor positioning system
  for labor tracking at construction sites: A case study in Guangzhou MTR}.
\newblock {\em Automation in Construction\/} {20}, 1 (2011), 3 -- 13.
\newblock
\showISSN{0926-5805}
\showDOI{%
\url{http://dx.doi.org/https://doi.org/10.1016/j.autcon.2010.07.009}}
\newblock
\shownote{Global convergence in construction.}


\bibitem{Xiao:2018:OMT:3190721.3190729}
{Fu Xiao}, {Zhongqin Wang}, {Ning Ye}, {Ruchuan Wang}, {and} {Xiang-Yang Li}.
  2018.
\newblock \showarticletitle{One More Tag Enables Fine-Grained RFID Localization
  and Tracking}.
\newblock {\em IEEE/ACM Trans. Netw.\/} {26}, 1 (Feb. 2018), 161--174.
\newblock
\showISSN{1063-6692}
\showDOI{%
\url{http://dx.doi.org/10.1109/TNET.2017.2766526}}


\bibitem{DBLP:journals/corr/abs-1709-01015}
{Faheem Zafari}, {Athanasios Gkelias}, {and} {Kin~K. Leung}. 2017.
\newblock \showarticletitle{A Survey of Indoor Localization Systems and
  Technologies}.
\newblock {\em CoRR\/}  {abs/1709.01015} (2017).
\newblock
\showURL{%
\url{http://arxiv.org/abs/1709.01015}}


\bibitem{zhuang2016smartphone}
{Yuan Zhuang}, {Jun Yang}, {You Li}, {Longning Qi}, {and} {Naser El-Sheimy}.
  2016.
\newblock \showarticletitle{Smartphone-based indoor localization with bluetooth
  low energy beacons}.
\newblock {\em Sensors\/} {16}, 5 (2016), 596.
\newblock


\end{thebibliography}

\end{document}